\documentclass[floatfix, nofootinbib, letterpaper, onecolumn, showpacs]{revtex4}
\usepackage{amssymb,amsthm,amsmath,float}

\pdfoutput=1
\usepackage[total={6.9in,9.3in}, top=0.8in, left=0.8in, includefoot]{geometry}
\usepackage{graphicx}

\usepackage{url}

\newcommand{\Eq}[1]{(\ref{eq:#1})}

\newcommand{\Lem}[1]{Lem.~\ref{lem:#1}}

\newcommand{\Sec}[1]{\S \ref{sec:#1}}
\newcommand{\Fig}[1]{Fig.~\ref{fig:#1}}

\newcommand{\InsertFig}[4]
{\begin{figure}[h!t]
 \centerline{
 \includegraphics[width=#4\columnwidth]{./figures/#1}
 }
 \caption{{\footnotesize #2}
 \label{fig:#3}}
\end{figure}}

\newcommand{\InsertFigTwo}[5] {
\begin{figure*}[h!t]
 \centerline{
 \includegraphics[width=#5\textwidth]{./figures/#1}
 \hskip 0.5in
 \includegraphics[width=#5\textwidth]{./figures/#2}
 }
 \caption{{\footnotesize #3}
 \label{fig:#4}}
\end{figure*}}

\newcommand{\bN}{{\mathbb{ N}}}

\newcommand{\bR}{{\mathbb{ R}}}
\newcommand{\bS}{{\mathbb{ S}}}
\newcommand{\bT}{{\mathbb{ T}}}
\newcommand{\bZ}{{\mathbb{ Z}}}
\newcommand{\cB}{{\cal B}}

\newcommand{\cS}{{\cal S}}
\newcommand{\cU}{{\cal U}}

\newcommand{\eps}{\varepsilon}

\newtheorem{thm}{Theorem}
\newtheorem{lem}[thm]{Lemma}

\newtheorem{prob}{Questions}

\newcommand{\beq}[1]{\begin{equation}\label{eq:#1}}
\newcommand{\eeq}{\end{equation}}

\newenvironment{se}[1]{\equation\label{eq:#1}\aligned}{\endaligned\endequation}
\newcommand{\bsplit}[1]{\begin{se}{#1}}
\newcommand{\esplit}{\end{se}}

\begin{document}
\title{Thirty Years of Turnstiles and Transport}
\author{J.~D.~Meiss}
\email{jdm@colorado.edu}
\affiliation{
	Department of Applied Mathematics\\
    University of Colorado \\
	Boulder, CO 80309-0526 \\
}
\date{\today}

\begin{abstract}
\vspace*{1ex}
\noindent
To characterize transport in a deterministic dynamical system is to compute exit time distributions from regions or transition time distributions between regions in phase space. This paper surveys the considerable progress on this problem over the past thirty years. Primary measures of transport for volume-preserving maps include the exiting and incoming fluxes to a region. For area-preserving maps, transport is impeded by curves formed from invariant manifolds that form partial barriers, e.g., stable and unstable manifolds bounding a resonance zone or cantori, the remnants of destroyed invariant tori. When the map is exact volume preserving, a Lagrangian differential form can be used to reduce the computation of fluxes to finding a difference between the action of certain key orbits, such as homoclinic orbits to a saddle or to a cantorus. Given a partition of phase space into regions bounded by partial barriers, a Markov tree model of transport explains key observations, such as the algebraic decay of exit and recurrence distributions. 

\end{abstract}

\maketitle

\section*{}
\textbf{The problem of transport in dynamical systems is to quantify the motion of collections of trajectories between regions of phase space with physical significance, for example, to determine chemical reaction rates, mixing rates in a fluid, or particle confinement times in an accelerator or fusion plasma device. For Hamiltonian systems with a phase space containing regular (elliptic islands and tori) and irregular (chaotic) components, transport is impeded by partial barriers bounding resonance zones or by remnants of invariant tori.
In the former case the barriers are formed from the broken separatrices of an unstable periodic orbit, and in the latter, they are formed from similar stable and unstable manifolds of a remnant torus---a cantorus.
Thirty years ago, MacKay, Meiss, and Percival discovered that cantori form robust partial barriers for area-preserving maps: in any region of phase space the most resistant barriers are the cantori. In this review, I discuss this work and survey subsequent developments. While much has been done, there are still many open questions.}

\section{Introduction}\label{sec:Introduction}

In this article, I attempt to survey the history of the theory of transport in deterministic, conservative dynamics, as well as to discuss some of the many questions that remain. My interest in this subject arose from a collaboration with Robert MacKay and Ian Percival that began just over thirty years ago with the publication of two papers on ``Transport in Hamiltonian Systems" \cite{MacKay84a, MacKay84}. Like many fruitful ideas, there was some synchronicity to this discovery: we learned at the ``Dynamics Days" meeting in June of 1983\footnote
{This was the sixth Dynamics Days and the fourth in Enschede, organized Robert Helleman.}
that David Bensimon and Leo Kadanoff were working on a similar approach \cite{Bensimon84}. 
I remember that when Robert, Ian, and I celebrated the acceptance of our papers, we thought that they contained ideas for at least ten years of additional research. Some thirty years later, I believe there are still many fruitful, open problems in this field. Much has been done: our articles have been cited 697 times, according to the Web of Science. The work of Vered Rom-Kedar and Stephen Wiggins, which introduced the term \textit{lobe dynamics} and (with Anthony Leonard) applied these ideas to fluid mixing, \cite{RomKedar90c, RomKedar90a} has also been extensively cited (443 times). 

The problem of deterministic transport also has been studied frequently in the first 24 volumes of \textit{Chaos: An Interdisciplinary Journal of Nonlinear Science}. Indeed the terms \textit{turnstile} or \textit{lobe} have been mentioned in 114 articles, beginning with an article on transport in a four-dimensional map in the very first issue \cite{Afanasiev91}. It would be impractical to try to review all of this work. Since I wrote a review in 1992 \cite{Meiss92}, I will concentrate here on describing some of the more recent advances, as well as on posing some questions that I hope will be answered before the $50^{th}$ anniversary of \textit{Chaos}. However, to make this paper coherent, it is necessary to tell some of the early story.

A theory of transport seeks to characterize the motion of groups of trajectories from one region of phase space to another. Typical dynamical systems have both regular and chaotic (irregular) trajectories. The question ``Where does this trajectory go?" is a particularly difficult one to answer for chaotic trajectories since they---by definition---exhibit sensitive dependence upon initial conditions.
The intermixture of regular and irregular dynamics is particularly evident in the Hamiltonian case where stable periodic orbits are surrounded by elliptic islands that are embedded in the chaotic sea as a fractal. 

So how does one study transport?
Our original paper \cite{MacKay84} set out a method:
\begin{quote}
In this paper we study transport in the irregular components and initiate a theory of the organization inherent in the apparent chaos. This is achieved by recognizing a natural division of the irregular components into regions separated by partial barriers formed by joining the gaps in invariant Cantor sets. We discuss the mechanism of transport between these regions. 
\end{quote}

There were many antecedents to these ideas. For example, in his paper introducing the quadratic area-preserving map, H\'enon computed the set of orbits that do not escape to infinity, a kind of anti-transport result \cite{Henon69}. More direct antecedents include the development of transition state theory in chemistry \cite{Pechukas79, Waalkens10, MacKay14} and especially the notions of Wigner for finding surfaces of minimal flux \cite{Wigner37, Keck67}. For a chemical reaction that has no activation energy and for which classical mechanics is a good approximation, Wigner thought there should be a surface in phase space that distinguishes between dissociated and reacted states. Unless the surface has ``wrinkles", that is, unless there are trajectories that cross it twice during one collision (similar to what MacKay later called ``sneaky returns" \cite{MacKay94a}), then the one-way flux across this surface gives a bound on the reaction rate.

For the purposes of this exposition consider a map $f: M \to M$ on some $n$-dimensional manifold $M$. Most of the ideas here can be applied directly to flows, either directly using a Poincar\'e or stroboscopic section, or by extension to continuous time \cite{MacKay86, MacKay94a, Mosovsky13}. For a map, a trajectory is a sequence $\{z_t\} = \{ \ldots, z_j, z_{j+1}, \ldots \}$, such that $z_t \in M$ and 
\[
	z_{t+1} = f(z_t).
\]
We will always assume that $f$ is a homeomorphism, and typically that it is a diffeomorphism.
In addition, we will suppose that $f$ preserves a volume measure $\mu$: 
for any measurable set $A$, 
\beq{volumePreserving}
	\mu(f(A)) = \mu(A) .
\eeq
When the map is a diffeomorphism, volume can be computed using a differential form
\beq{VolumeForm}
	\Omega = \rho(z) dz^1\wedge dz^2\wedge \ldots \wedge dz^n ,
\eeq
with density $\rho$, i.e., $\mu(A) = \int_{A} \Omega$. Then $f$ preserves volume when
\beq{VP}
	f^*\Omega = \Omega .
\eeq
Here the \textit{pullback}, $f^*$, \label{page:pullback} is the local action of the map $f$ on differentials. An easy mnemonic for how this works is to denote the image as a function $z'(z) = f(z)$, and then $f^* dz = dz'$ is the differential of this function, i.e., $f^* dz^i = \sum_{j} Df_{i,j}(z) dz^j$. Here $Df_{i,j} = \partial f^i / \partial z^j$ is the Jacobian matrix of $f$. The implication is that \Eq{VP} for the form \Eq{VolumeForm} implies that
\[
	\det({Df}(z)) \rho(f(z)) = \rho(z) .
\]
For the common case that $\rho(z) = 1$, this implies that the Jacobian of $f$ has determinant one.

\section{Area-Preserving Maps}\label{sec:APMaps}
Most of the theory of transport has been developed for the area-preserving case. For an \textit{area-preserving map}, there are coordinates $z = (x,y)$ such that the preserved volume form is $\Omega = dy \wedge dx$.\footnote
{It is convenient to choose this orientation to give conventionally oriented boundary integrals.
Note that that this means clockwise orientation is positive.}
The well-studied Chirikov standard map on the cylinder $M = \bS \times \bR$,
\beq{StdMap}
	(x',y') = \left(x+y' \mod 1,\, y-\tfrac{k}{2\pi} \sin(2\pi x)\right) ,
\eeq
is one such family. A typical phase portrait for this map is shown in \Fig{StdMapLeaking}.
The standard map is an example of a \textit{twist} map on the cylinder: a map for which
the image of a vertical line is a graph over the angle \cite{Meiss92}. The powerful Aubry-Mather theory applies to this class of maps \cite{Mather82, Aubry83a}. Another example is H\'enon's family of area-preserving maps,
\beq{Henon}
		(x',y') = (-y+2(a-x^{2}), y) .
\eeq

\InsertFig{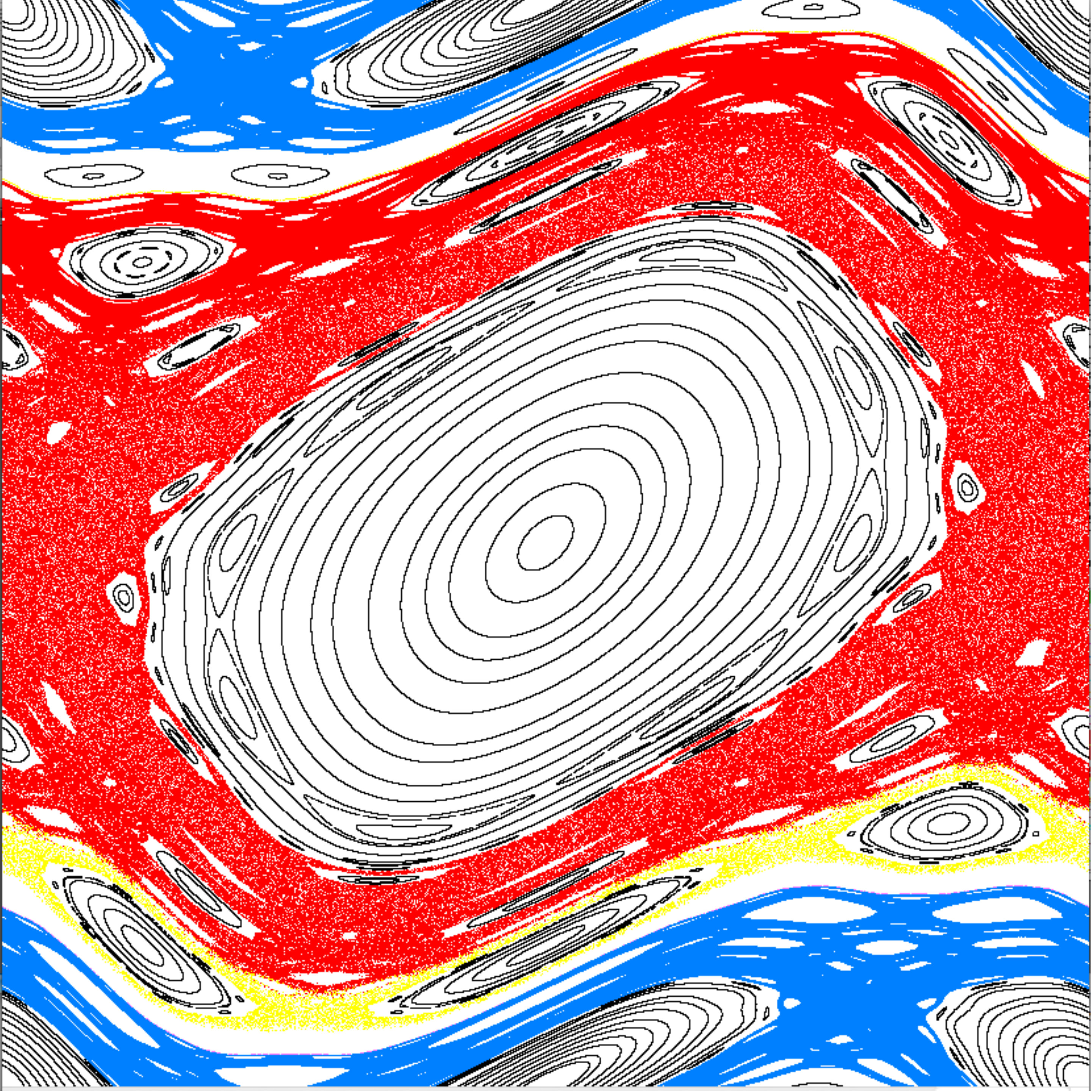}
{Phase portrait of the standard map \Eq{StdMap} for $k = 0.975$, on the domain $[-0.5,0.5] \times [-0.5,0.5]$. Black trajectories show island chains associated with various elliptic periodic orbits. There are no rotational invariant circles for this value of $k$.
The red trajectory, started near the hyperbolic fixed point at $(0.5,0.0)$, is 
iterated $10^6$ times and appears to be confined by rotational circles; however, it will escape
after $2.3\times10^9$ steps. The blue and yellow trajectories are similarly only temporarily confined. This figure was generated with the \textit{StdMap} application \cite{Meiss08}.}{StdMapLeaking}{0.7}

Area-preserving maps are a special case of symplectic maps on a $2d$-dimensional manifold that preserve a closed two-form. 
Symplectic maps are of interest because they arise as canonical Poincar\'e maps for Hamiltonian flows \cite{Meiss92}. Thus the theory of \cite{MacKay84} applies to the calculation of transport in a two degree-of-freedom Hamiltonian system and has applications to mechanics, chemical reactions \cite{Pechukas79, Waalkens10, MacKay14}, plasma confinement \cite{Hazeltine03, Boozer04}, celestial mechanics \cite{Wisdom91,Koon00, Murray01}, particle accelerator design \cite{Wilson01}, and condensed matter \cite{Toller85}. More generally, volume-preserving maps arise as maps between sections or as stroboscopic maps for incompressible flows. Most physical applications correspond to three-dimensional flows such as the motion of a passive scalar in a fluid \cite{Cartwright99a, Wiggins04a, Haller15} or mixing in granular media \cite{Sturman08}. 
There are also nontrivial examples of periodically time-dependent two-dimensional flows, such as Aref's blinking vortex \cite{Aref84} for which the area-preserving theory is directly applicable.

\subsection{Periodic Orbits and Resonance Zones}\label{sec:Resonances}
Periodic orbits for area-preserving maps are typically born in pairs, one a saddle (hyperbolic) and the other elliptic. For example, the standard map \Eq{StdMap} for $k = 0$ becomes
the trivial shear $(x,y) \mapsto (x+y,y)$; it integrable with action variable $y$. For this map, each circle on which $y$ is rational is composed of period-$n$ orbits since 
\begin{align*}
	(x_n,y_n) &= f^n(x_0,y_0) \\
	          &= (x_0 + ny_0 \mod 1 \,,\, y_0) = (x_0,y_0) ,
\end{align*}
when $y_0 = \frac{m}{n}$ with $m\in \bZ$ and $n \in \bN$. The minimal period of the orbit is $n$ if $\gcd(m,n) = 1$. These periodic orbits then have rotation number $\omega = \frac{m}{n}$, and we call them $(m,n)$-orbits.

When the integrable shear is perturbed, at least two orbits survive for each rational rotation number; these are called the \textit{Birkhoff orbits}, since they are predicted by the famous Poincar\'e-Birkhoff theorem \cite{Meiss92}. In \Fig{StdMapLeaking} such orbits can be seen at the centers of island chains. For the case of a \textit{twist map}, like \Eq{StdMap}, Aubry-Mather theory shows that these orbits always exist; they are the so-called \textit{minimizing} and \textit{minimax} orbits because they can be obtained from a variational principle based on the action \cite{Mather82, Aubry83a}. The minimizing orbit is hyperbolic whenever it is a nondegenerate minimum of the action, and the minimax orbit is elliptic when the perturbation is small enough \cite{MacKay83b}.

An elliptic-hyperbolic pair of $(m,n)$-orbits forms the skeleton of a \textit{resonance zone} consisting of $n$ islands surrounding the elliptic points with boundaries formed from the stable and unstable manifolds of the saddle \cite{MacKay87, Easton91}. One can view an island chain as a region of phase space in which the local rotation number is, roughly speaking, $\frac{m}{n}$. Computational methods for delineating resonances include Laskar's \textit{frequency map} \cite{Laskar93a} and the \textit{width} of an orbit relative to a given rotation number \cite{Easton93}.

\InsertFig{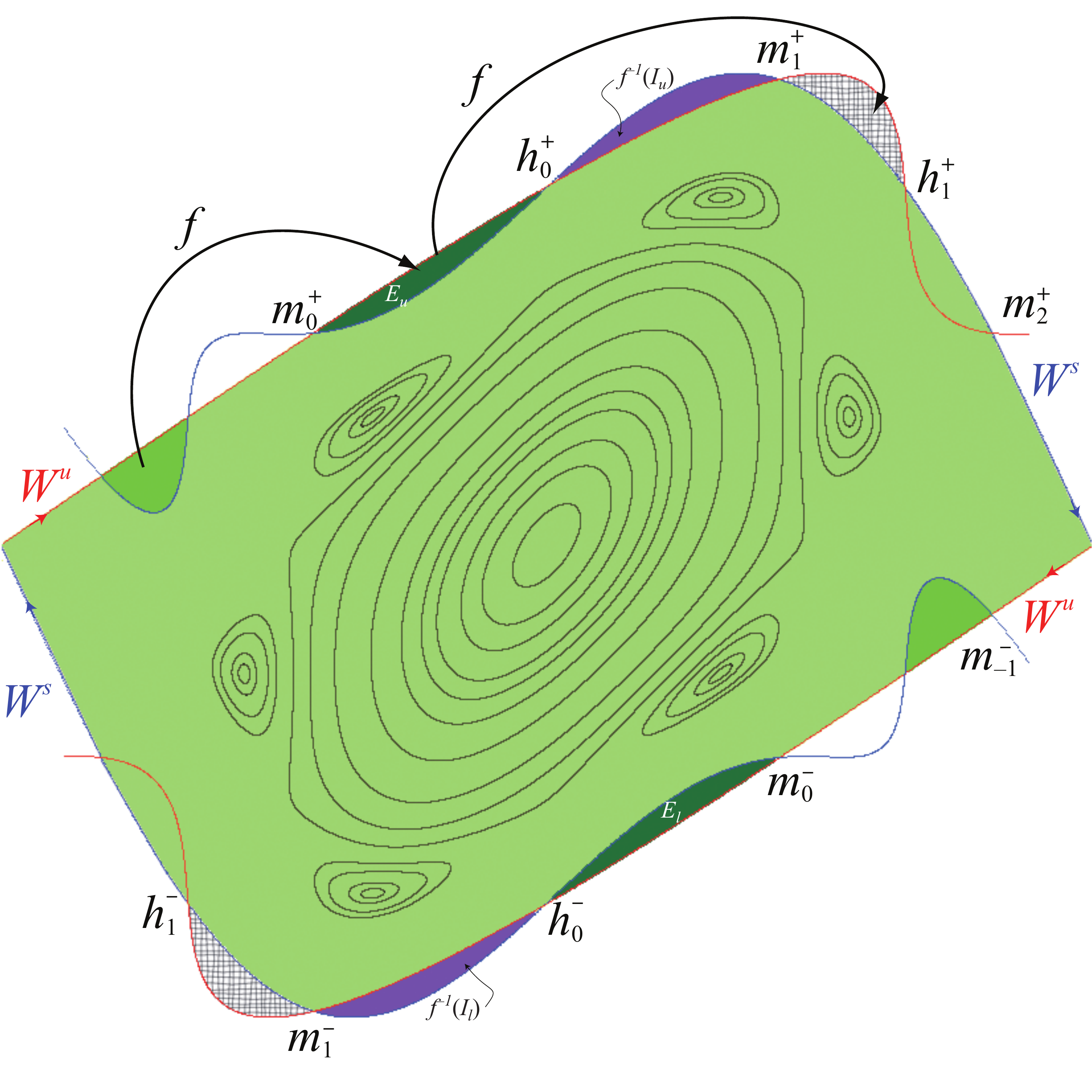}{Resonance zone (shaded region) for the $(0,1)$ orbits of \Eq{StdMap} for $k=1.5$. The boundary of the zone consists of two segments of unstable (red) and two segments of stable (blue) manifold of the $(0,1)$ hyperbolic point joined at the primary homoclinic points $m_1^\pm$. The upper and lower exit sets (dark green) and their first image (gray hatching) and preimage (light green) are shown, as are the preimage of the incoming sets (violet).}{01Resonance}{0.6}

Geometrically, one can define a region of phase space that is bounded by broken separatrices of the saddle, an example is shown in \Fig{01Resonance}.
The boundary is formed using points on primary homoclinic orbits. Recall that a homoclinic point,
\[
	m \in W^u(h)\cap W^s(h),
\]
is \textit{primary} if there are segments of stable, $W^s(h)$, and unstable, $W^u(h)$, manifold from the saddle $h$ to $m$ (that is, \textit{initial} segments) that intersect only at their endpoints \cite{Easton86, RomKedar90c} (see \cite{Lomeli00a} for a generalization to higher dimensions).
The resonance boundary in \Fig{01Resonance} consists of two initial segments of unstable manifold---for the ``upper" and ``lower" boundary---connected to two initial segments of stable manifold at primary homoclinic points. The region enclosed by the pair of broken separatrices is the resonance zone. The switch from unstable to stable manifold can occur at any point on the homoclinic orbit $\{m_t\}$; even though the resulting geometry depends on this choice, the area of the resonance does not.
In order for an orbit to leave the resonance zone it must pass through either the upper, $E_u$ or lower, $E_l$, \textit{exit set} (we will say more about these in \Sec{Flux}). Similarly, an orbit can enter the zone only by first landing in the preimage of one of the incoming sets, $f^{-1}(I_u)$ or $f^{-1}(I_l)$.
This construction can be done for any pair of $(m,n)$-orbits; a $(1,3)$ resonance zone for \Eq{StdMap} is shown in \Fig{13Resonance}. 

Each exit and incoming set is a \textit{lobe} for the resonance, and the set $E_u \cup f^{-1}(I_u)$, is a \textit{turnstile}: it acts like a rotating door.\footnote
{The name ``turnstile" was suggested to us by Carl Murray.}
For each iteration of the map, trajectories escape through $E_u$ and arrive from $f^{-1}(I_u)$. 

\InsertFig{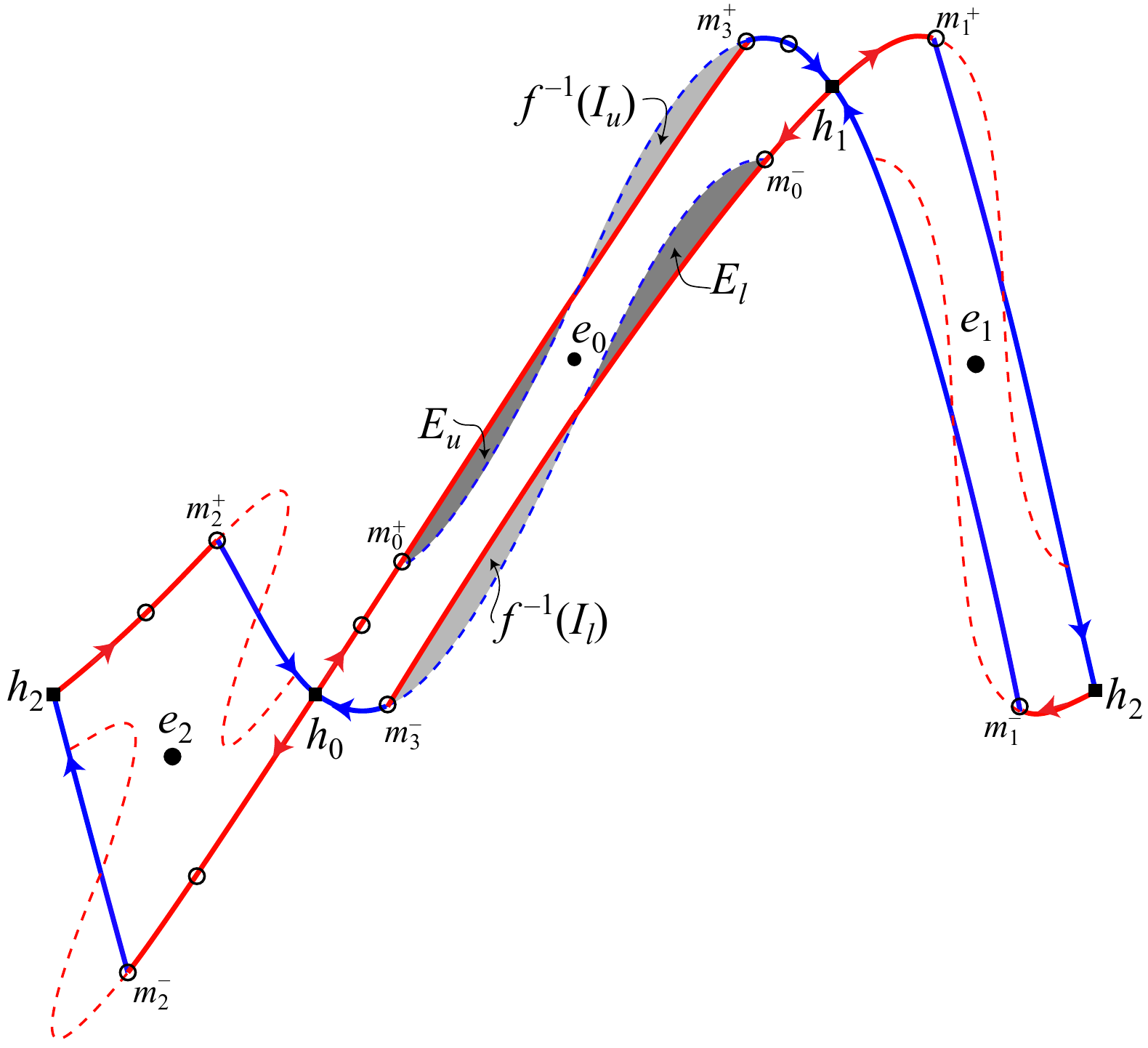}{Resonance zone for the $(1,3)$-orbits of the standard map when $k=1.5$.
The elliptic orbit (solid circles) is labeled $\{e_0,e_1,e_2\}$, the hyperbolic (solid squares) $\{h_0,h_1,h_2\}$. The $\circ$ symbols label a pair of (minimizing) homoclinic orbits $\{m_t^{\pm}\}$ on the unstable (red) and stable (blue) manifolds of the saddle. The switch from unstable to stable manifold island occurs at the points $m_3^\pm$ and their two preimages. The shaded regions are turnstiles composed of an exit set, $E_u$ ($E_l$) (dark grey) and the preimage of an incoming set $I_u$ ($I_l$) (light grey).}
{13Resonance}{0.6}

\subsection{Cantori}\label{sec:Cantori}
The existence of invariant tori with sufficiently irrational frequency vectors in Hamiltonian systems that are sufficiently close to integrable is guaranteed by KAM theory \cite{Poshel01}. However, KAM theory says nothing about what happens to these structures far from integrability. The existence of remnants of these tori after they have been destroyed was first deduced by Serge Aubry and Ian Percival using variational arguments \cite{Aubry78, Percival79}. Percival named the remnants \textit{cantori} because---for an area-preserving map---they are invariant Cantor sets. Every orbit on a cantorus has the same irrational rotation number. 

Figure~\ref{fig:Cantori} shows the invariant circle of \Eq{StdMap} with golden mean rotation number, $\omega = \phi$, at the threshold of its destruction, $k = k_{cr}(\phi) \approx 0.97163540631$ \cite{Greene79}, and at seven larger values of $k$ where it has become a cantorus.
For twist maps, the projection of a cantorus onto the angle coordinate, $x$, is a Cantor set on the circle, so it has a countably infinite set of gaps. The orbit of each gap is a \textit{family} of related gaps. For the standard map each cantorus appears to have one family, and each gap grows continuously from zero as $k$ increases from $k_{cr}(\omega)$. More generally, there may be more than one family of gaps and invariant circles may reform \cite{Baesens93}.

\InsertFig{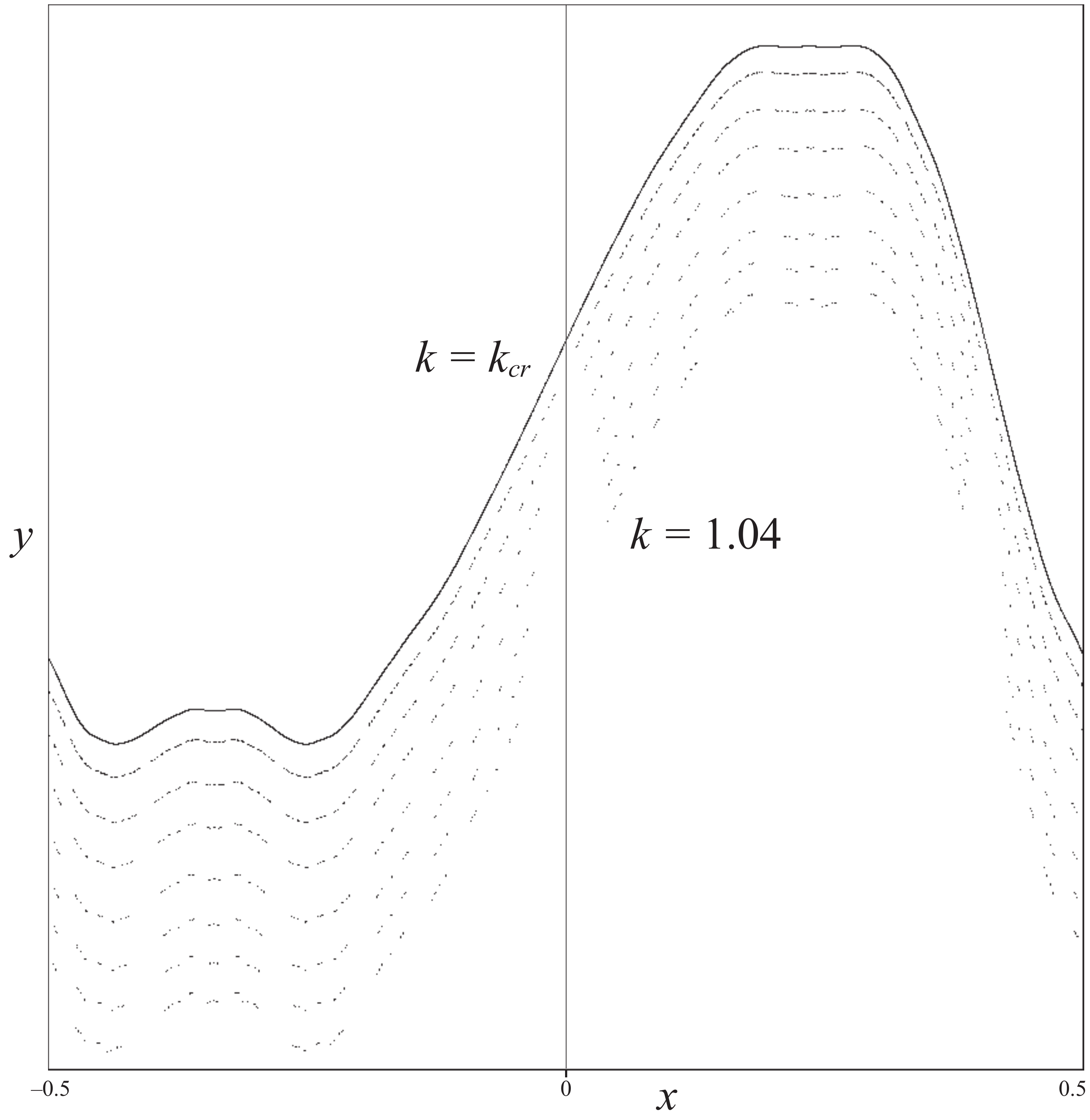}{Destruction of the golden mean invariant circle and creation of a cantorus for the standard map. At $k= k_{cr}(\phi)$ there is an invariant circle (top curve), but it is not smooth. Successive cantori, for $k = 0.98$ up to $1.04$, are displaced downward for clarity. The largest gap forms around $x=0$. The displayed orbits are actually saddle periodic orbits with rational rotation numbers that are convergents to the golden mean, see \Sec{CantoriBarriers}.}{Cantori}{0.7}

Mather and Aubry showed that for twist maps there is an invariant set for each irrational rotation number that is either an invariant circle or a cantorus \cite{Mather82, Aubry83a}. The same analysis applies to optical Hamiltonian flows \cite{Gole01}. It is not clear, however, whether maps that do not satisfy a twist condition have cantori. Cantori do exist for a large class of maps close enough to an anti-integrable limit \cite{Aubry90, Aubry95}.

\begin{prob}[Nontwist Cantori]
What is the fate of invariant tori for area-preserving maps that do not satisfy a twist condition? Are there ``shearless" cantori for nontwist maps like those studied by \cite{Fuchss06}?
\end{prob}

The anti-integrable limit also can be used to show that certain higher-dimensional symplectic maps \cite{MacKay92b} and higher-order recurrence relations modeling crystal lattices have analogues of cantori \cite{Knibbler14}. However, Aubry-Mather theory has not been generalized to these cases.

\begin{prob}[Higher-Dimensional Cantori] When an invariant torus in a multi-dimensional symplectic map is destroyed, does it immediately become a cantorus, or are there other topological configurations such as Sierpinski curves? The mechanism for destruction of two-tori in volume-preserving maps seems analogous to the symplectic case \cite{Fox13a}, but are there remnant cantori? 
\end{prob}


\section{The Problem of Transport}\label{sec:Transport}

The inherent loss of predictability in chaotic dynamical systems implies that computation of individual trajectories is not practical, nor is it especially useful. A theory of transport seeks to make predictions about the longtime evolution of ensembles of trajectories and answer questions such as ``How long does it take, on average, to get from one region to another?", or ``What is the probability distribution of escape times from a region?" Thus for the chemical reaction $AB + C \to A + BC$, the first region might correspond to the separated reactants $AB$ and $C$, and the second to the products $A$ and $BC$. In plasma physics the first region may correspond to the interior of a confinement device such, as a tokamak, and escape to the divertor.

The first step in a theory of transport is the computation of the \textit{flux} across codimension-one surfaces. For chaotic maps, this idea was used by Channon and Lebowitz \cite{Channon80} who computed escape times for H\'enon's quadratic map. They used a broken separatrix of a period-five orbit to construct a simple, closed curve through which orbits can escape. 
In this section, I review the notions of flux and net flux for volume-preserving and exact volume-preserving maps.

It is important to note that one of the fundamental measures of chaos, the Lyapunov exponent is not usually a good measure for transport. In \cite{MacKay84} we said:

\begin{quote}
The Lyapunov exponents give the rate at which a nice cat in a region of phase space turns into a mixed-up cat. The turnstile rate constants give the mean rate at which bits of the cat are transported to regions of the phase space...
\end{quote}

\subsection{Exit Sets and Flux}\label{sec:Flux}
A first step in a global theory of transport is to quantify the motion of an ensemble for short times by computing the flux of trajectories across codimension-one surfaces in phase space. Suppose $R$ is a region with a piecewise smooth boundary $\partial R$.
When $R$ is invariant, $f(R) = R$, and $\partial R$, is a ``complete" barrier. When $R$ is ``almost invariant" its boundary is a \textit{partial barrier}: though trajectories may escape, the leakage is slow. For a flow, the flux through a surface is computed by integrating the normal component of the velocity over the surface \cite{MacKay86, MacKay94a}. For a map, the flux is the volume that escapes from the region upon each iterate, i.e., the volume of the exit set $E \subset R$---the subset whose first image is not in $R$:
\bsplit{ExitSet}
	 E 	&= \{ z \in R \,|\, f(z) \notin R\} 
	 	= \{ z \in R \,|\, z \notin f^{-1}(R)\} \\
	 	&\equiv R \setminus f^{-1}(R) ,
\esplit
see \Fig{Flux}(a).
The flux of escaping orbits from $R$ is just the volume of the exit set,
\beq{Flux}
	\Phi(R) = \mu(E) = \int_E \Omega .
\eeq
Similarly, the incoming set $I$ for a region $R$ is the set of points that have just entered
\[
	I = R \setminus f(R) . 
\]

Formally, $R$ is almost invariant in the sense of measure if $\Phi(R) \ll \mu(R)$. One technique to compute almost invariant sets uses the Perron-Frobenius operator; this idea was pioneered by Dellnitz and Junge \cite{Dellnitz99}. In lieu of this method, we will construct regions using resonance zones and cantori, for which the flux can be localized to turnstiles, see \Sec{Action}.

\InsertFigTwo{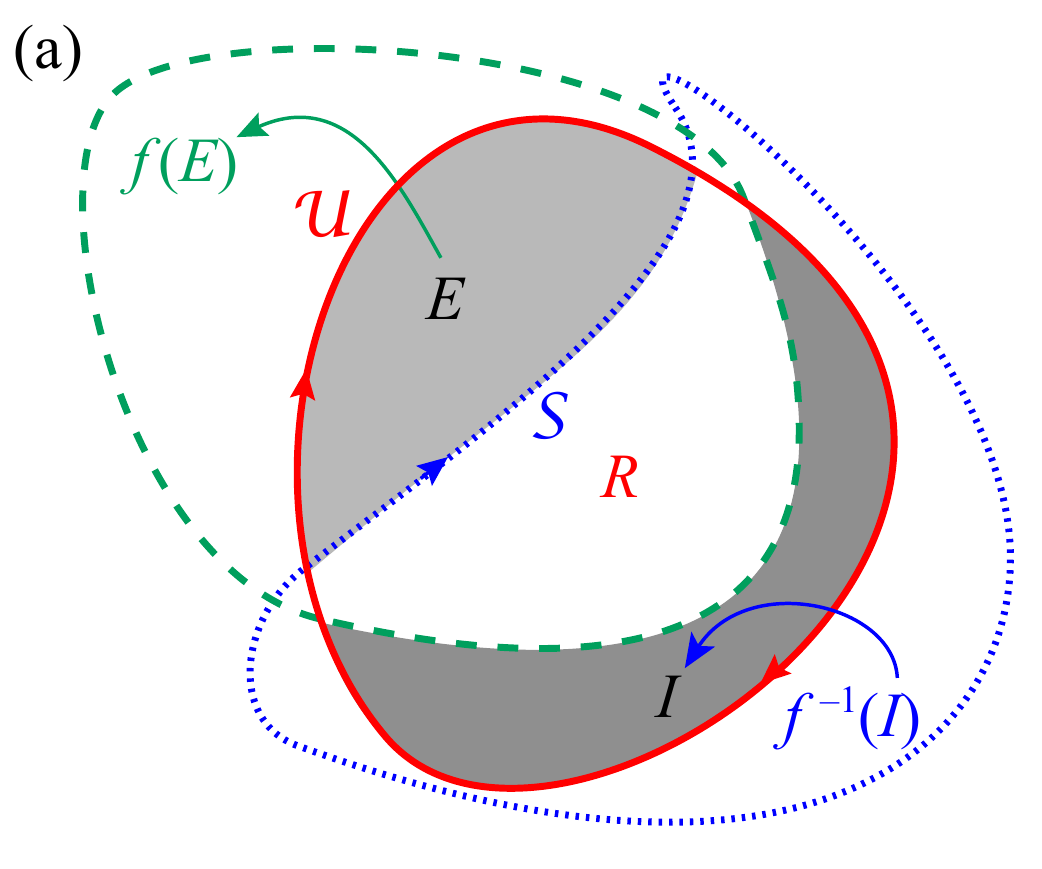}{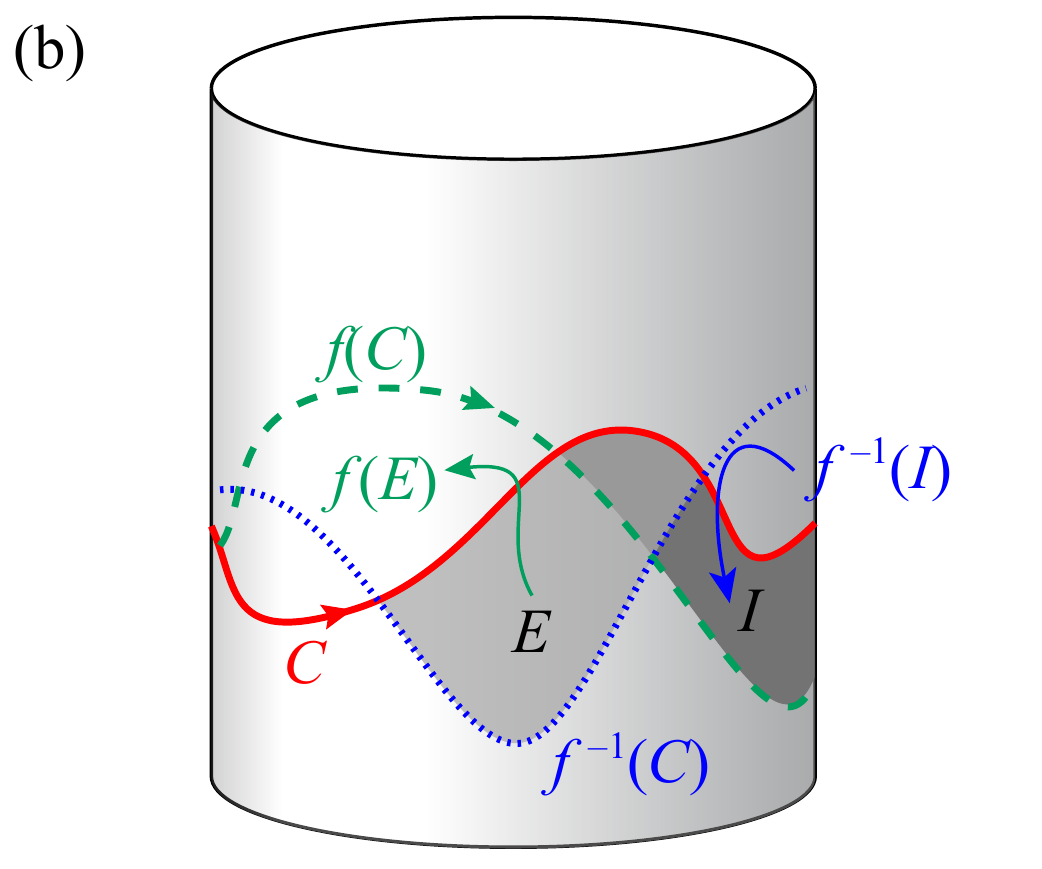}{Exit and Incoming Sets. (a) A region $R$, its exit set $E$ with boundary $\cU-\cS$, and incoming set $I$. Orientations of the boundaries are shown by the arrows. (b) A rotational circle $C$ on the cylinder and its image (dashed) and preimage (dotted). The net flux vanishes when the region above $C$ and below $f(C)$, labeled $f(E)$ has the same area as that below $C$ and above $f(C)$, labeled $I$.
}{Flux}{0.45}

A volume form \Eq{VolumeForm} $\Omega$ is exact if
\beq{OmegaExact}
	\Omega = d\alpha
\eeq
for some $(n-1)$-form $\alpha$. For example, the standard map \Eq{StdMap} is volume preserving with $\Omega = dy \wedge dx$. Since $x \in \bS^1$ is periodic, we cannot integrate $\Omega$ with respect to this variable, but the \textit{Liouville form} $\alpha = y dx$, is well-defined on the cylinder and satisfies \Eq{OmegaExact}. When \Eq{OmegaExact} holds, Stokes's theorem can be used to reduce volume integrals to surface integrals.

The boundary of the exit set, $\partial E$, is the union of two oriented subsets $\cU \subset \partial R$ and $\cS \subset f^{-1}(\partial R)$ as sketched in \Fig{Flux}(a). For a resonance zone, these are pieces of unstable and stable manifolds, recall \Fig{01Resonance}.
Assuming that the boundaries are chosen with an orientation consistent with that of $R$, then $\partial E = \cU - \cS$, and we can use \Eq{OmegaExact} to rewrite \Eq{Flux} as
\beq{BoundaryFlux}
	\Phi(R) = \mu(E) = \int_{E} d\alpha = \int_{\partial E} \alpha 
			= \int_{\cU} \alpha - \int_{\cS}\alpha \;.
\eeq
For two-dimensional maps, $\cU$ and $\cS$ are curves, and the line integrals in \Eq{BoundaryFlux} are more efficient to compute than the original two-dimensional integral over $E$.

In a sense, the complete transport problem is solved if one knows the orbits of $E$ and $I$ for each region of interest. For example, the minimum transit time from $A$ to $B$ is the first time that $f^t(E_A) \cap I_B \neq \emptyset$. Similarly, a complete description of transport through a region requires knowing only the future history of the entering trajectories, i.e., the sets $f^t(I)$. These ideas form the basis for the theory of \textit{lobe dynamics} as formulated by Rom-Kedar and Wiggins \cite{RomKedar90c, RomKedar94}. As we will recall in \Sec{ExitTime}, this is especially useful in the study of transport through resonance zones. 

\subsection{Exact Maps and Net Flux}\label{sec:Exact}

When a region $R$ has finite measure, it is easy to see that the \textit{net flux} through $\partial R$ vanishes, so that the measure of the incoming and exit sets must be the same: 
\begin{align*}
	\mu(I) &= \mu(R) - \mu(R \cap f(R)) 
	 		= \mu(R) - \mu(f^{-1}(R) \cap R ) \\
	 		&= \mu(R) - \mu(R \cap f^{-1}(R)) 
	 		= \mu(E) ,
\end{align*}
where we used \Eq{volumePreserving} and \Eq{ExitSet}.

However, when the phase space is not contractible there may exist closed surfaces that that do not bound a region. For example the standard map \Eq{StdMap} models a kicked rotor for which the canonical coordinate is an angle, and the phase space is the cylinder $\bS \times \bR$. A \textit{rotational circle} is a simple closed curve that is not contractible, for example $C_{0}= \{(x,0): x \in \bS\}$. An area-preserving map on the cylinder need not have zero net flux through rotational circles; a simple example is the map $f(x,y) = (x,y+1)$ that shifts each rotational circle upward but still has $\det{Df} = 1$. 

Zero net flux is equivalent to exactness: a map $f$ is \textit{exact} volume-preserving when there exists a \textit{Lagrangian form} $\lambda$ such that
\beq{Exact}
	f^*\alpha - \alpha = d\lambda .
\eeq
where $\alpha$ satisfies \Eq{OmegaExact} (recall the discussion of pullback on page \pageref{page:pullback}).
The Lagrangian for a two-dimensional map is a zero-form---a function---and is the discrete-time analog of the continuous-time Lagrangian of mechanics. Indeed for a flow, $\lambda$ is the time-integral of the Lagrangian along a trajectory between sections \cite{MacKay86, Wendlandt97}. 

For example using $\alpha = y dx$, the standard map \Eq{StdMap} is exact with the Lagrangian
\beq{StdMapLagrangian}
	\lambda(x,y) = \tfrac12 y'^2 + \tfrac{k}{4\pi^2} \cos(2\pi x) ,
\eeq
where $y'(x,y) = y-\tfrac{k}{2\pi} \sin(2\pi x)$. This follows because
\begin{align*}
	f^*\alpha - \alpha &= y'dx' - ydx = y'd(x + y') - y dx \\
	 			 &= y' dy' + (y-\tfrac{k}{2\pi} \sin(2\pi x))dx - y dx \\
	 			&= d( \tfrac12 y'^2 + \tfrac{k}{4\pi^2} \cos(2\pi x)) ,
\end{align*} 
the differential of \Eq{StdMapLagrangian}. In our original work for area-preserving twist maps we wrote $\lambda = F(x,x')$, which is one of the standard generating functions for canonical transformations \cite{MacKay84}. 
I first learned of the more general formulation from Bob Easton \cite{Easton91}.

For $n$-dimensional maps $\lambda$ is an $(n-2)$-form. For example, the ABC map \cite{Feingold88},
\beq{ABCMap}
	\begin{pmatrix} x' \\ y' \\z'\\ \end{pmatrix}
	 = \begin{pmatrix}	x + A \sin z + C \cos y \\
 						y + B \sin x' + A \cos z\\
 						z + C \sin y' + B \cos x'
 \end{pmatrix} ,
\eeq
preserves the standard volume $\Omega = dx \wedge dy \wedge dz$, a fact that is most easily seen by viewing it as the composition of three shears, $f = s_z \circ s_y \circ s_x$, where, $s_x(x,y,z) = (x+ A\sin z+C\cos y, y,z)$, etc. Though it is natural to view \Eq{ABCMap} as a map on the three-torus, it can also be thought of as a map on the cover $M = \bT^2 \times \bR$. Now the volume form $\Omega$ is exact with $\alpha = z dx \wedge dy$, and \Eq{ABCMap} is also exact with Lagrangian one-form
\begin{align*}
	\lambda =& A\left[ \sin(z)(dx'+zdy) + \cos(z)(dy-zdx')\right] \\
	        &+ B\sin(x') dy' + C\cos(y')dx' .
\end{align*}
Here the primed variables are to be thought of as the functions defined in \Eq{ABCMap}. Such a representation is especially pertinent when $B,C \ll 1$, since then $z$ is an action-like variable (invariant when $B=C=0$), and one can use it to compute fluxes that correspond to vertical motion.

It is also possible to generalize the idea of canonical generating functions to \textit{generating forms} for volume-preserving transformations when a generalized twist condition is satisfied \cite{Lomeli09a}. 

\begin{prob}[Generating Forms] Can Lagrangian forms be used for perturbation theory and near-integrable normal forms, just as canonical generators are used for the symplectic case? Can they be used to build volume-preserving integrators, like those found for area-preserving maps in \cite{Wendlandt97}? 
\end{prob}

When $f$ is homotopic to the identity, the net-flux ``across" a closed $(n-1)$-dimensional surface $S$ has meaning: it is the signed volume contained in the region $R$ bounded by $S$ and $f(S)$. If $\partial R = f(S)-S$, Stokes's theorem implies that
\bsplit{ZeroNetFlux}
	\Phi_{Net}(S) &= \int_R \Omega = \int_{f(S)-S} \alpha \\
	              &= \int_S (f^*\alpha-\alpha) = \int_S d\lambda = 0 ,
\esplit
because the integral of a total differential (exact form) over a closed surface $S$ vanishes. For example in \Fig{Flux}(b), the rotational circle $C$ on the cylinder is the upper boundary of an unbounded set with  $y \to -\infty$. This set has exit set $E$, the region below $C$ whose image is above $C$, and incoming set $I$, the region below $C$ whose preimage is above $C$. The region contained between $f(C)$ and $C$ is $R = f(E) \cup I$, and when the map has zero net flux \Eq{ZeroNetFlux} implies that $\mu(f(E)) = \mu(E) = \mu(I)$.

Maps with nonzero net flux also have interesting transport properties \cite{Fox15}.
\section{Action and Flux}\label{sec:Action}
Perhaps the most surprising aspect of the calculation of \cite{MacKay84} is the connection between Mather's action difference and the flux of trajectories across curves in phase space. 

\InsertFig{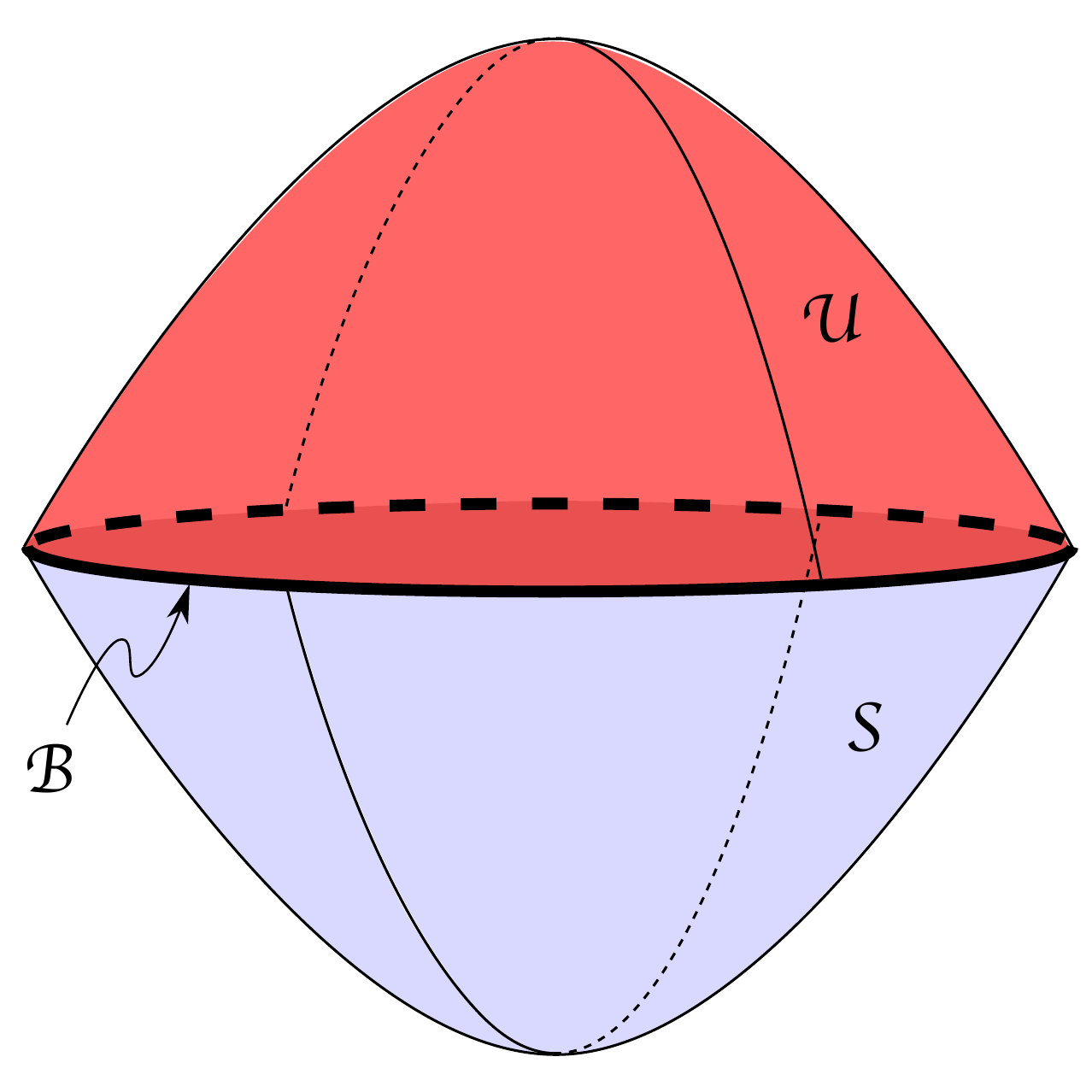}{A three-dimensional lobe bounded by a pair of surfaces $\cU$ (dark red) and $\cS$ (light blue) that intersect along a loop $\cB$ on their common boundaries.}{3DLobe}{0.4}

In the general context of an exact volume-preserving map, \Eq{Exact}, let $\cS$ and $\cU$ be a pair of codimension one surfaces that together bound a region $E$, as sketched in \Fig{3DLobe}. Suppose that
\[
	\cB = \cU \cap \cS = \partial \cU = \partial \cS 
\]
is the common $(n-2)$-dimensional boundary of the surfaces.
The volume of $E$ is given by the difference between integrals over the surfaces as in \Eq{BoundaryFlux}. Each of these $(n-1)$-dimensional integrals can be partially integrated using \Eq{Exact} with Stokes's theorem; for example,
\beq{Iterative}
	\int_\cS \alpha = \int_\cS (- d\lambda + f^*\alpha ) 
					= - \int_{\cB} \lambda + \int_{f(\cS)} \alpha .
\eeq
This iterative formula can be used to compute lobe volumes in a number of different contexts. For example, suppose that $\cU = f^{n}(\cS)$; then we can iterate \Eq{Iterative} $n$-times to obtain
\beq{NSteps}
	\int_\cS \alpha = -\sum_{t=0}^{n-1} \int_{f^{t}(B)} \lambda + \int_{f^n(\cS)} \alpha .
\eeq
Note that the final integral in \Eq{NSteps} is, by assumption, exactly the integral over the surface $\cU$, and so from \Eq{BoundaryFlux} we have
\beq{PeriodicAction}
	\mu(E) = \int_\cU \alpha - \int_\cS \alpha = \sum_{t=0}^{n-1} \int_\cB f^{*t}\lambda .
\eeq
The last sum represents a generalized \textit{action} of the periodic boundary $\cB = f^n(\cB)$.

A second case in which \Eq{Iterative} is useful is when $\cS$ is a patch of stable manifold of some lower-dimensional invariant set, i.e., $\cS \subset W^s(A)$, where $A$ is a normally hyperbolic invariant manifold. In this case the limit of the remainder integral in \Eq{NSteps} as $n \to \infty$ is zero, and thus 
\beq{ForwardAsymptotic}
	\int_\cS \alpha = -\sum_{t=0}^{\infty} \int_{f^{t}(B)} \lambda 
\eeq
is the forward action sum. Similarly, if $\cU \subset W^u(B)$ is a patch of unstable manifold of an invariant set $B$, it is appropriate to reverse \Eq{Iterative} to iterate backwards, and obtain
\beq{BackwardAsymptotic}
	\int_\cU \alpha = \sum_{t=-\infty}^{-1} \int_{f^{t}(B)} \lambda .
\eeq

Lomel\'i and I showed how to compute flux for a $3D$ volume-preserving map \cite{Lomeli09b}, and MacKay showed how to compute flux over a saddle in a Hamiltonian flow \cite{Mackay89b,MacKay14} (a result implicit in \cite{Toller85}, as well). Nevertheless, the general problem in higher dimensions still has many mysteries.

\begin{prob}[Multidimensional Flux]
Can one use the flux formulas to compute flux through $2D$ manifolds in a $3D$ map, see e.g., \cite{Mireles13}, and through resonance zones formed from $2D$ normally hyperbolic invariant sets in a $4D$ symplectic map, see e.g., \cite{Gillilan91}?
\end{prob}

\subsection{Flux Through Periodic Orbits}\label{sec:Periodic Flux}
When $n=2$, $\cB$ in \Eq{Iterative} degenerates to a set of points, and $\int_\cB \lambda$ reduces to evaluation at its upper and lower limits. For example, suppose that $\cU$ and $\cS$ are curve segments joined at their endpoints, points on a pair of period-$n$ orbits. An example would be e.g., the pair $\{e_0\}$ and $\{h_0\}$ that form an island chain, recall \Fig{13Resonance}. 
In this case $\lambda$ is a function, and the boundary $\cB = e_0 - h_0 $ is a pair of points, $e_0$ on the minimax orbit and $h_0$ on the hyperbolic orbit. Then \Eq{PeriodicAction} becomes
\bsplit{PeriodicFlux}
	\mu(E) = \sum_{t=0}^{n-1} \left[\lambda(e_t)- \lambda(h_t) \right] = \Delta W[e,h] &,\\
	\mbox{(Periodic Orbit Flux)}&,
\esplit
the difference between the actions of the two orbits.

Indeed, $\Delta W[e,h]$ is the flux ``through" the pair of orbits, as can be seen by constructing (if possible) a region $R$ with boundary $C$ that satisfies two hypotheses.
\begin{itemize}
	\item[(H1).] Suppose $C = \partial R$ is a simple closed curve alternately joining successive points on a pair of period-$n$ orbits, $\{e_t\}$ and $\{h_t\}$.
	\item[(H2).] Suppose that $f(C) \cap C = \{e_t\} \cup \{h_t\}$. 
\end{itemize}
An example of such a curve is sketched in \Fig{OrbitFlux}. The requirement (H2) means that the exit and incoming lobes are bounded by segments of $C$ and $f(C)$ that connect pairs of points on the two orbits. By definition \Eq{ExitSet}, $f(E)$ is the region outside $C$ and inside $f(C)$. Each of the $n$-lobes of the image of the exit set is bounded by $\cU_{t} -\cS_t$ where $\cS_t$ ($\cU_t$) is the segment of $C$ ($f(C)$) from $h_t$ to $e_t$. Since $f(\cS_t) = \cU_{t+1}$, and $f(\cS_{n-1}) = \cU_0$, \Eq{Exact} implies that the escaping flux is
\begin{align*}
	\Phi(R) &= \mu(f(E)) = \sum_{t=0}^{n-1} \int_{\cU_t-\cS_t } \alpha 
					 = \sum_{t=0}^{n-1} \int_{\cS_t} f^*\alpha - \alpha \\
			 		&= \sum_{t=0}^{n-1} \int_{\cS_t} d\lambda 
					= \sum_{t=0}^{n-1}(\lambda(e_t) -\lambda(h_t)) = \Delta W[e,h].
\end{align*}
Thus the flux crossing $C$ is exactly the action difference of \Eq{PeriodicFlux}. It is especially interesting, I think, that the flux is independent of the choice of curve $C$, given that it satisfies (H1) and (H2). Such curves are easy to construct for twist map examples, where the $(m,n)$ orbits are ordered relative to $x$, but it is not obvious that it is always possible to construct such curves for a pair of period $n$ orbits.

\InsertFig{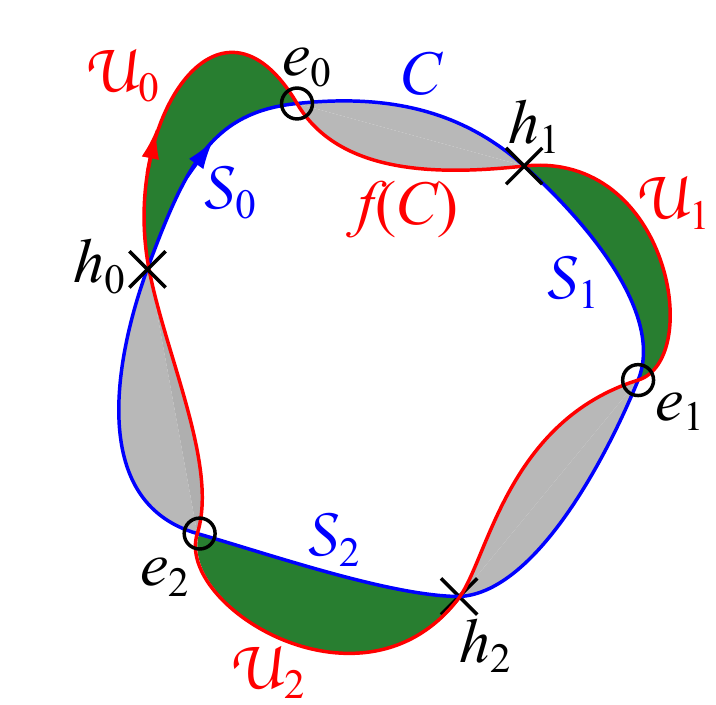}{Sketch of the construction of a partial barrier through a pair of $(1,3)$ orbits. The exited area is the region outside $C$ and inside $f(C)$, the dark (green) shaded areas. Under (H1) and (H2), this flux is \Eq{PeriodicFlux}. }{OrbitFlux}{0.4}

\begin{prob}[Periodic Flux] Under what conditions is it possible to find a curve $C$ that satisfies (H1) and (H2)?
\end{prob}

\subsection{Separatrix Lobe and Resonance Areas}\label{sec:ResonanceArea}
The iterative formula \Eq{Iterative} can also be used to compute lobe and resonance zone areas for $2D$ maps. Recall from \Sec{Resonances} that a resonance zone is bounded by segments of stable and unstable manifold of a hyperbolic periodic orbit joined at a primary homoclinic point. A separatrix lobe is a region bounded by segments of stable and unstable manifold that intersect on a neighboring pair of primary homoclinic points.

More generally, following \cite{Easton91}, suppose that $R$ is a region bounded by $K$ alternating segments of stable and unstable manifold, thus $\partial R = \sum_{j=1}^{K}(\cU^j + \cS^j)$ where the lengths of $f^{t}(\cS^j)$ and $f^{-t}(\cU^j)$ limit to zero as $t \to \infty$ since they are stable and unstable segments, respectively. 
Denote the $2K$ ``corner points", where these curves are joined, by $m^j$ and $h^j$, $j = 1, 2, \ldots K$, so that $\partial \cU^j = m^j-h^j$ and $\partial \cS^j = h^{j+1}-m^j$, with $h^{K+1} = h^1$. Labeling the orbits of the corner points with subscripts, $f^t(m^j) = m_t^j$, as usual, then \Eq{ForwardAsymptotic} and \Eq{BackwardAsymptotic} become 
\begin{align*}
	\int_{\cS^j} \alpha &= -\sum_{t=0}^{\infty} \left(\lambda(h_t^{j+1}) - \lambda(m_t^j)\right) ,\\
	\int_{\cU^j} \alpha &= \sum_{t=-\infty}^{-1} \left(\lambda(m_t^j) -\lambda(h_t^j)\right) .
\end{align*}
Finally, summing over the $K$ alternating segments to obtain the full boundary of $R$ implies,
from \Eq{BoundaryFlux},
\beq{MMPAction}
	\mu(R) = \int_{\partial R} \alpha = \sum_{t=-\infty}^\infty \sum_{j=1}^{K} 
			\left(\lambda(m_t^j) - \lambda(h_t^j) \right) .
\eeq

For example suppose that $f$ and $p$ are saddle fixed points, and that $\cU \subset W^u(f)$ and $\cS \subset W^s(p)$ are segments that intersect on a pair of heteroclinic points $h_0$ and $m_0$ as shown in \Fig{2DLobe}. Then \Eq{MMPAction} applies with $K=1$: there are just two segments and two corner points. Thus the area of the lobe is the action difference
\bsplit{LobeArea}
	\mu(R) = \sum_{t=-\infty}^\infty \left(\lambda(m_t) - \lambda(h_t) \right) 
		   = \Delta W[m,h] &,\\ 
	       \quad \mbox{(Heteroclinic Lobe Area)}&.
\esplit
This applies to the computation of the lobe area shown in \Fig{01Resonance}. The same formula applies when the saddles are periodic orbits.

\InsertFig{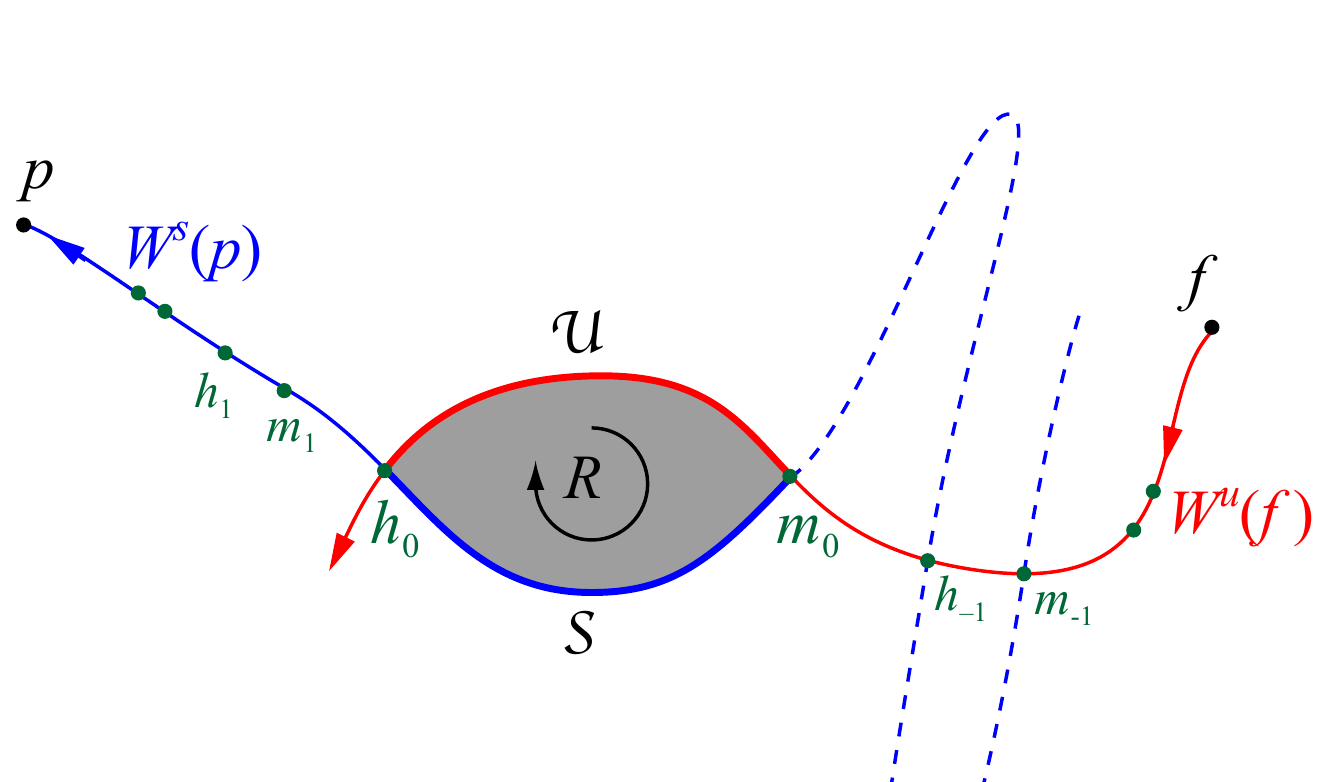}{Lobe formed from segments of stable, $\cS$ and unstable, $\cU$, manifold
of hyperbolic fixed points $p$ and $f$, respectively, with heteroclinic orbits $\{h_t\}$ and $\{m_t\}$. The lobe area $\mu(R)$ is the difference in between the actions of heteroclinic orbits, \Eq{LobeArea}.}{2DLobe}{0.6}

The action principle can also be used to compute the area of a resonance zone as well, recall \Sec{Resonances}. 
Each island in the resonance is bounded by four segments of manifold and so the result \Eq{MMPAction} applies with $K=2$. There are four corner points: $h^l$ and $h^r$, the ``left" and ``right" saddle points of one island, and $m^\pm$, the ``upper" and ``lower" homoclinic points. For example, for the period-three resonance of \Fig{13Resonance}, the leftmost island has corners $h^l = h_2$, $m^+_2$, $h^r= h_0$, and $m^-_2$. Since each of the islands have the same area, the total area of the resonance is $n$ times that of one island:
\bsplit{ResonanceArea}
	\mu(R) = n \sum_{t=-\infty}^\infty \left (\lambda(m^+_t) - \lambda(h^l_t) 
				+ \lambda(m^-_t) -\lambda(h^r_t)\right)&,\\
	 \quad \mbox{(Period-$n$ Resonance Area)} &.
\esplit
The result was first used to compute resonance areas in \cite{MacKay87}. Explicit results were obtained for the sawtooth map, a piecewise linear version of \Eq{StdMap} \cite{Chen89, Chen90}.

For a twist map like \Eq{StdMap}, the rotational periodic orbits are vertically ordered according to their rotation numbers, and there is a resonance zone for each $(m,n)$-minimizing/minimax pair of periodic orbits. A choice of homoclinic points can be made so that different zones intersect only in their turnstiles, and since the net flux vanishes, the areas of the two intersections must be equal. Thus one can partition phase space into rotational resonance zones \cite{MacKay87}. An example is shown in \Fig{ResonancePartition}. Qi Chen showed that when there are no rotational invariant circles, the total area of the infinitely many resonance zones is the full area of phase space, that is, the partition is ``complete" \cite{Chen87}.
One way to visualize this is to plot the area below each rotational curve as a function of rotation number \cite{MacKay87}. Rotational curves can be formed from invariant circles, cantori, or broken separatrices. This area function jumps by the value \Eq{ResonanceArea} at every rational rotation number. The partition is a complete \textit{devil's staircase} when the total rise of the staircase is the sum of the jumps.

\begin{prob}[Resonance Partition] Is the resonance partition complete for nontwist area-preserving maps when there are no rotational invariant circles? Is there a similar (complete) partition for one-action, volume-preserving maps? How could one construct a resonance partition for $4D$ symplectic maps (some of the intriguing geometry has been recently explored in \cite{Lange14})?
\end{prob}

\InsertFig{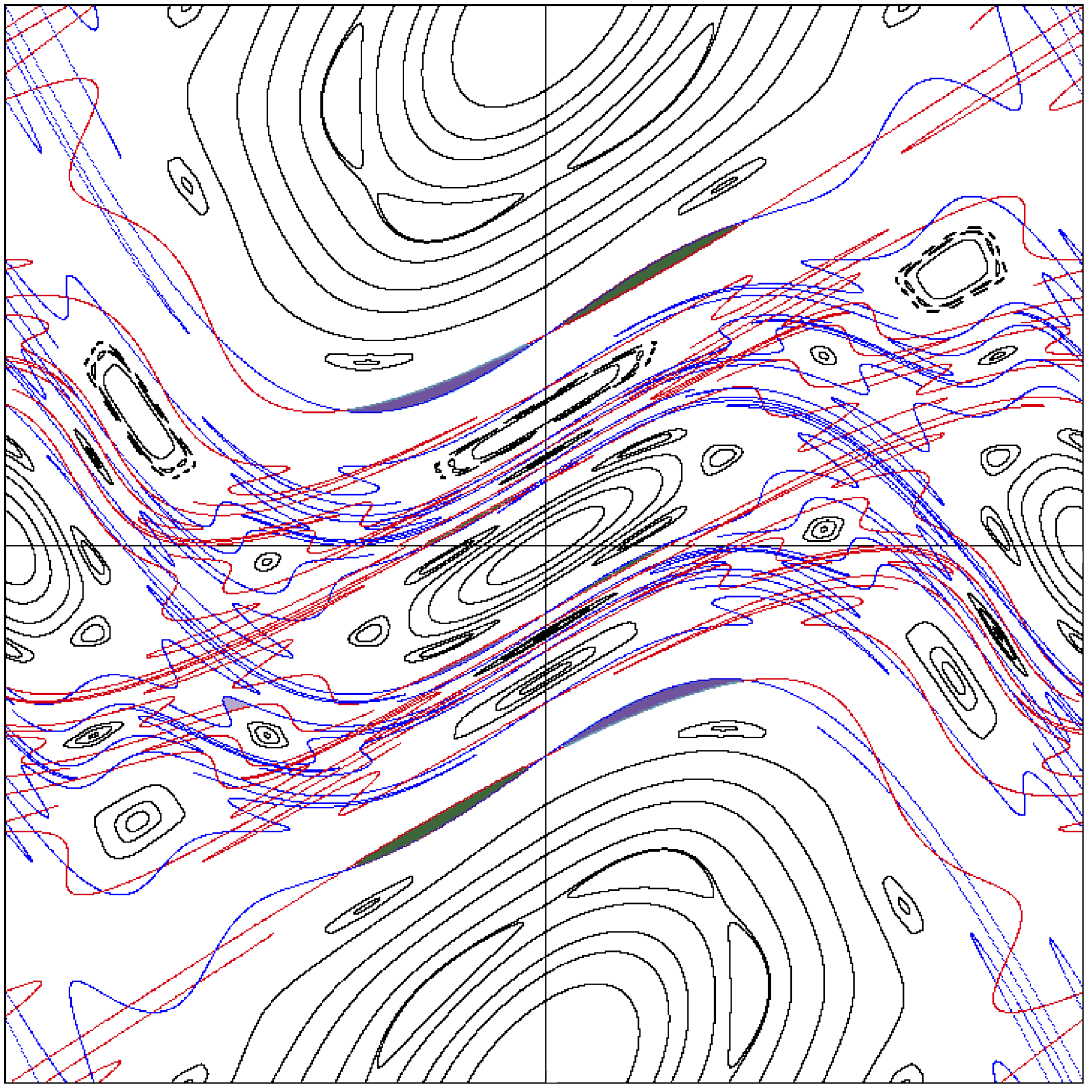}
{Partition of the phase space of the standard map into resonance 
zones for $k = 1.1$. Shown are portions of the stable (blue) and unstable (red) manifolds for the $\tfrac01$, 
$\tfrac13$, $\tfrac12$, $\tfrac23$, and $\tfrac11$ resonance zones. Exit sets (green) and 
Incoming sets (violet) for the $\tfrac01$ and $\tfrac11$ resonances shown. The partition is complete because there are no rotational invariant circles.}
{ResonancePartition}{0.7}

\subsection{Cantori as Partial Barriers}\label{sec:CantoriBarriers}

During the early years of the study of chaos, it was commonly observed that transport in a mixed phase space is highly nonuniform and is particularly slow as an orbit wends its way through regions containing recently destroyed tori. Indeed, Chirikov's famous numerical experiments for the standard map seemed to show that the time to cross from $y \approx 0$ to $y\approx 0.5$ goes to infinity as $k \downarrow 0.989$ \cite{Chirikov79a}; recall \Fig{StdMapLeaking}. Thus it seemed reasonable to speculate that cantori, even though they are typically hyperbolic, could act as strong barriers to transport. The importance of cantori was confirmed in \cite{MacKay84} when we discovered that the flux is locally minimal on cantori. We also were able to explain Chirikov's result, showing that the average crossing time for \Eq{StdMap} has a power law singularity with exponent $3.05$ at $k_{cr}(\phi) \approx 0.971$. 

For symplectic twist maps, every orbit on an invariant Lagrangian torus is a minimum of the action \cite{MacKay89c}. For the area-preserving case, as noted in \Sec{Cantori}, a destroyed invariant circle becomes a cantorus: the minimizing orbits persist and are typically hyperbolic. Suppose that a cantorus $T_\omega$ with rotation number $\omega$ has is a single family of gaps: there are two minimizing orbits $\{l_t\}, \{r_t\} \subset T_\omega$ that correspond to the left and right endpoints of the gaps. When the cantorus is hyperbolic there exist $c,\lambda >0$ such that $\|r_t -l_t\| < c e^{-\lambda |t|}$. Aubry-Mather theory implies that there is another distinguished orbit $\{m_t\}$, the minimax orbit, that falls in the gaps of the cantorus and is homoclinic to the orbit of gap endpoints, i.e., $m_t \to l_t$ as $t \to \pm \infty$.

The flux through $T_\omega$ can be computed using the general action principle \Eq{MMPAction}. Suppose $C$ is any curve that satisfies hypotheses (H1) and (H2): it is a simple closed curve through the cantorus that crosses each gap going through the minimax point, and $f(C) \cap C = T_\omega \cup \{m_t\}$ is exactly the cantorus and the minimax orbit. The same analysis as before implies that the flux crossing $C$ is the action difference
\bsplit{CantorusFlux}
	\Phi(T_\omega) = \sum_{t=-\infty}^\infty \left(\lambda(m_t) -\lambda(l_t)\right)
	               = \Delta W[m,l] &,\\ 
	\mbox{(Cantorus Flux)} &.
\esplit
This action difference is also prominent in Aubry-Mather theory: when it is zero $T_\omega$ is an invariant circle, and when it is positive, $T_\omega$ is a cantorus \cite{Mather82, Aubry83a}. 

Computing invariant tori is difficult since orbits with irrational rotation numbers are infinitely long. One algorithm uses a Newton-like method to compute the Fourier series for the conjugacy to rigid rotation of an invariant torus \cite{Huguet12}. However, this method does not seem to permit the computation of cantori. Another algorithm approximates a torus by periodic orbits. For the $2D$ case, this can be done using the continued fraction expansion $\omega = [a_0,a_1,\ldots]$, with $a_0 \in \bZ$, $a_j \in \bN$, of the rotation number to give a sequence of rationals
\bsplit{Convergents}
	\frac{m_j}{n_j} &= [a_0,a_1,a_2,\ldots a_j] \\
	                &\equiv a_0 + 1/(a_1 + 1/(a_2 + \ldots + 1/a_j)),
\esplit
the \textit{convergents} of $\omega$.
A similar technique uses the Farey tree, which, beginning with an interval $[\tfrac{m_l}{n_l}, \tfrac{m_r}{n_r}]$ bounded by \textit{neighboring} rationals, $m_r n_l - m_l n_r = 1$, constructs the daughter $m_d = m_l + m_r$, and $n_d = n_l + n_r$, with $\tfrac{m_l}{n_l} < \tfrac{m_d}{n_d} <\tfrac{m_r}{n_r}$. Repeating this construction on the two new intervals generates a binary tree that contains every rational in the original interval, and for which every irrational is a limit of an infinite path.

The flux through a cantorus can be approximated as the flux through a sequence of approximating $(m,n)$-orbit pairs, using \Eq{PeriodicFlux}. An example of the resulting flux function on the Farey tree is shown in \Fig{FluxFareyTree} for \Eq{StdMap} at $k=k_{cr}(\phi)$ when the golden mean invariant circle is at the threshold of destruction. What is remarkable is that this function appears to monotonically decrease: the flux through a Farey daughter is always smaller than that through each of its parents. For a Farey path that limits to the rotation number of an invariant circle, the flux goes to zero; this is the case in \Fig{FluxFareyTree} for $\omega = \phi^{-2}$, which has the path
\[
	\frac12,\, \frac 13,\, \frac25,\, \frac38,\, \frac{5}{13},\, 
	\frac{8}{21},\, \frac{13}{34} ,\, \ldots .
\]
When the Farey path limits to the rotation number of a cantorus, the periodic flux \Eq{PeriodicFlux} limits to \Eq{CantorusFlux}. Thus the strongest barriers, corresponding to the smallest flux, are associated with cantori. 

An alternative notion of minimal flux through cantori was formulated by Polterovich \cite{Polterovich88b}.

\InsertFig{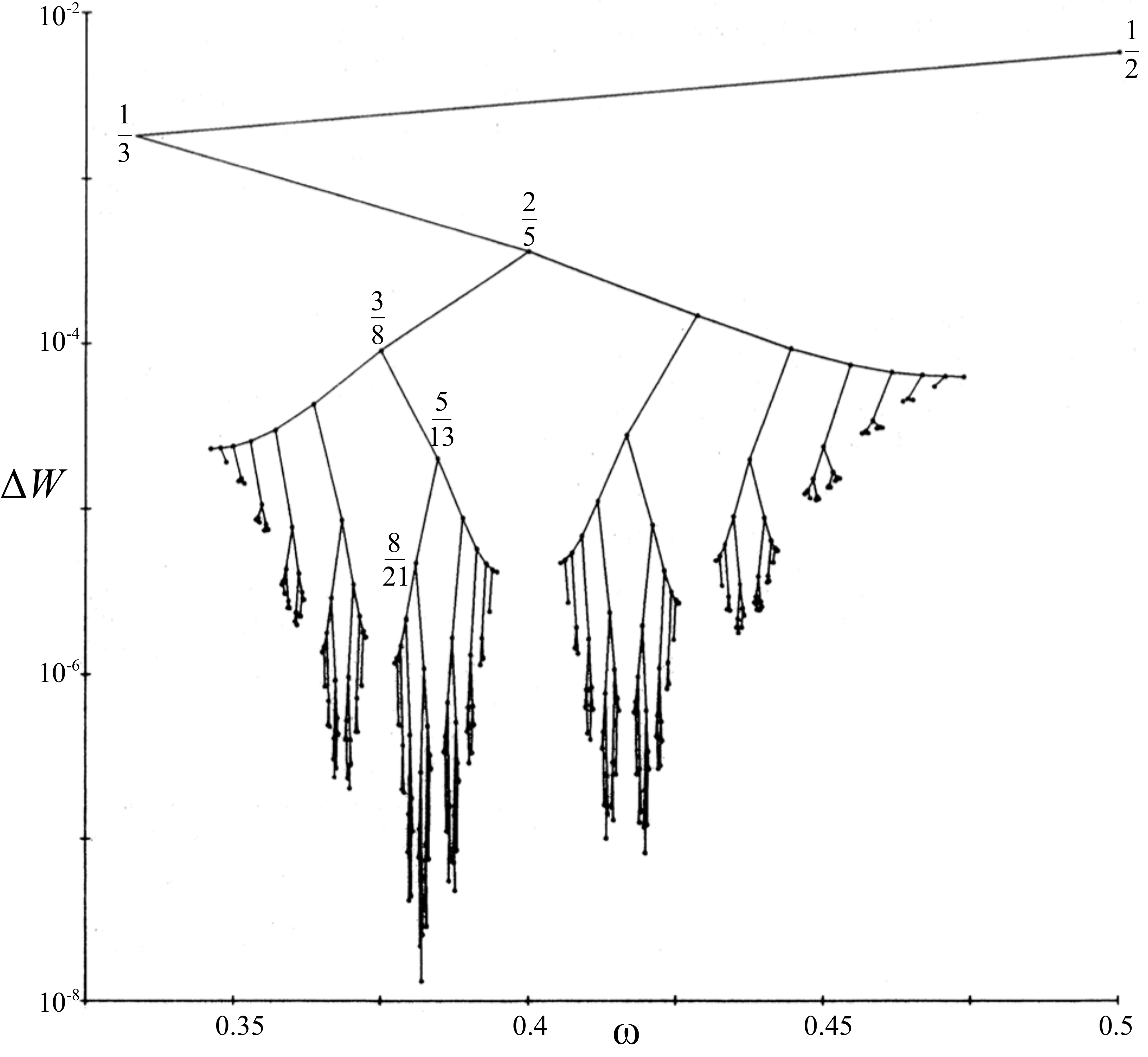}{Flux for the standard map at $k=k_{cr}(\phi)$ through periodic orbits on the Farey Tree. Reproduced with permission from Physica D 20:387 (1986) Copyright 1986 Elsevier Publishing.}{FluxFareyTree}{0.7}

\begin{prob}[Cantori and Minimal Flux] Find a category of twist maps for which the flux is provably a monotone decreasing function on the Farey tree. For example, if the map is analytic, is the flux asymptotically monotone?
\end{prob}

Though there is no completely satisfactory generalization of continued fractions to higher dimensions, Kim and Ostlund invented a binary tree that has most of the properties of the single frequency Farey tree \cite{Kim86} (it does not have a unique path to each rational vector). Even though there have been many numerical studies of the break-up of invariant tori for symplectic maps \cite{Kook89, Tompaidis96a, Celleti04b} and volume-preserving maps \cite{Fox13a}, there is no conclusive evidence for the formation of remnant tori that are Cantor sets.

\begin{prob}[Higher-Dimensional Cantori] Is there a multidimensional generalization of the Farey tree or continued fraction that is optimal for approximating tori and cantori? After destruction, is a torus replaced by a remnant invariant set with the same rotation vector; if so, what is its topology?
\end{prob}

\section{Exit and Return Times}\label{sec:ExitTime}

Though a region's escaping and entering fluxes give one measure of transport, fluxes represent only the first step, both dynamically and theoretically. More global measures are obtained from distributions of exit, crossing, and return times \cite{RomKedar90c,Meiss97}. Early numerical experiments were done by Channon and Lebowitz, who showed that the exit time distribution from a resonance zone for the H\'enon map \Eq{Henon} has a surprising power law form \cite{Channon80}. This observation was greatly extended by Karney \cite{Karney83}, who used integer arithmetic to eliminate floating point errors, and also by Chirikov and Shepelyansky \cite{Chirikov84a}. From the many subsequent observations of power law decays, it seems clear that this phenomena is due to the presence of stable structures within the region of interest: these structures are extremely \textit{sticky}. However, there is still much controversy about whether the asymptotic decay is truly a power law, and---if it is---if there are universal values for the exponent.

\subsection{Exit Time Distributions}\label{sec:Distributions}

Given a subset $A$ of phase space $M$, the \textit{(forward) exit time} of a point $a \in A$ is the number of iterates for it to leave:
\beq{ExitTime}
	t^+_A(a) \equiv \min_{t>0} \{t: f^t(a) \notin A\} .
\eeq
Similarly, the \textit{backward exit time} is
\[
	t^-_A(a) \equiv \min_{t>0} \{t : f^{-t}(a) \notin A\} .
\]
In either case, the exit time is set to $\infty$ if the images remain in $A$ for all $t \in \bN$. By definition \Eq{ExitSet}, the exit set, $E$, of a region is the set with forward exit time one, and the incoming set, $I$, is the set with backward exit time one. The \textit{transit time} is the time for an orbit to move from the incoming to the exit set of $A$; since this is constant along an orbit we can define
\beq{TransitTime}
	t^{transit}_A(a) = t^+_A(a) + t^-_A(a) -1 ,
\eeq
subtracting the $1$, to measure the time to get from $I$ to $E$.
Finally, the \textit{return time} for a point $a \in A$ is
\beq{CrossingT}
	t^{return}_{A}(a) \equiv \min_{t> 0}\{ t: f^t(a) \in A\} .
\eeq
Note that if $a$ is not in the exit set, its return time is $1$ since the minimum is taken over positive $t$. 

Chaos by its very definition precludes accurate long-time computations of  \Eq{ExitTime}-\Eq{CrossingT}. Nevertheless, one would like to obtain estimates of their distributions. Whenever $A$ is has finite, nonzero measure, its \textit{exit time distribution},
\beq{ExitDistribution}
	p^\pm_A(j) = \frac{\mu(\{a \in A: t^\pm_A(a) = j\})}{\mu(A)} ,
\eeq
is the probability of exiting in time $j$. 
Another common measure of transport is the distribution of recurrence times: the probability 
that an orbit beginning in $A$ has first return time $j$,
\[
	p^{return}_A(j) = \frac{\mu(\{a \in A: t^{return}_A(a) = j\})}{\mu(A)} .
\]
It is not hard to relate this distribution to others that we have defined. If a point in $A$ is  not in the exit set, then its return time is $1$, so we have
\[
	p^{return}_A(1) = \frac{\mu(A \setminus E)}{\mu(A)}= 1- \frac{\mu(E)}{\mu(A)}.
\]
Points that exit $A$ are in $E$, and $f(E)$ is the incoming set for $M\setminus A$. The first return to $A$ thus occurs when the orbit transits $M\setminus A$, so
\beq{ReturnDistribution}
	p^{return}_A( j) = \frac{\mu(E)}{\mu(A)} p^{transit}_{M \setminus A}(j-1), \quad j>1 .
\eeq
Thus the return time distribution to $A$ is essentially the transit time distribution through $M \setminus A$.

To compute the statistics of exit and return times, any invariant subset of $A$ should be excluded since it has infinite exit time. For example, the islands of stability that generically surround any elliptic periodic orbit should be excluded from exit time computations for the resonance zone of an area-preserving map. It is convenient to define the \textit{accessible subset} of $A$ is the subset of points that can be reached from the outside, or equivalently, the set of points with finite backward exit time:
\[
	A_{acc} \equiv \{a \in A : t^-(a) < \infty\} .
\]
This set has the same volume as the set with finite forward exit time. In order that the distribution \Eq{ExitDistribution} be normalized over $\bN$, it should be redefined using $A_{acc}$ instead of $A$.

In 1947 Mark Kac proved a result for stochastic processes that can be translated into a result for deterministic maps \cite{MacKay84, MacKay94a, Meiss97}. Namely, suppose that $M$ has finite measure and the invariant measure is normalized so that $\mu(M) = 1$. If $M_{acc}$ is the subset of $M$ that is accessible to orbits beginning in $A$, then
the average first return time to $A$ is
\beq{AverageReturn}
	\langle t^{return}_A \rangle_A \equiv \frac{1}{\mu(A)} \int_A t^{return}_A d\mu
							 = \frac{\mu(M_{acc})}{\mu(A)} .
\eeq
Using the relation between transit time and return time distributions \Eq{ReturnDistribution}, gives the equivalent result:

\begin{lem}[Average Exit Time]\label{lem:ExitTime}
 If $A$ is a measurable set with incoming set $I$ and accessible set $A_{acc}$ the average exit time for points in the incoming set is
\beq{AverageExit}
	\langle t^+ \rangle_I = \langle t^{transit}_A \rangle_{I} 
						 = \frac{ \mu (A_{acc})}{\mu(I)} .
\eeq
\end{lem}

\noindent
One interesting aspect of \Lem{ExitTime} is that it implies that the exit time distribution has finite mean, and by \Eq{ReturnDistribution}, that the mean return time to $A$ is also finite when $\mu(M) = 1$. Higher moments of these distributions need not be finite; a counterexample is given by a simple shear map \cite{Meiss97}.

Since \Eq{ReturnDistribution} is normalizable then one can define its cumulative distribution, the probability that the return is at least as long as $k$, 
\[
	P^{return}_A(k) = \sum_{j=k}^{\infty} p^{return}_A(j).
\]
This is often called the \textit{Poincar\'e recurrence distribution} \cite{Zaslavsky85, Chirikov84a, Chirikov99}. The Poincar\'e recurrence theorem states that when $\mu(M) = 1$, almost all trajectories return, i.e., $P^{return}_A(1) = 1$.

In most cases there are no analytical formulae for exit and return time distributions. For a small perturbation to an integrable map with a homoclinic connection to a saddle, however, Melnikov theory can be used to estimate the exit set volume \cite{RomKedar90a, Wiggins92}. Melnikov theory, originally developed for flows, was formulated for maps by Easton \cite{Easton84,Lomeli08a}. Another case that can be treated is that of an adiabatic perturbation, where the lobes cover the region swept by the separatrices of the frozen time subsystems \cite{Kaper91}.
If the structures of the homoclinic tangle or trellis can be given a topological classification, like that developed by Easton \cite{Easton86, Easton98}, then these methods can be extended to approximate additional intersection areas, and hence the exit time distributions \cite{RomKedar94, Mitchell06, Mitchell09}.

\begin{prob}[Higher-Dimensional Trellises] Can one classify trellises in $3D$ volume-preserving and $4D$ symplectic maps to obtain information about transport properties?
\end{prob}

\subsection{Stickiness and Anomalous Diffusion}\label{sec:Stickiness}

If a map is uniformly hyperbolic then the number of periodic orbits of a given period grows exponentially at a rate given by the topological entropy. This results in an exponential form for exit time distributions \cite{Zaslavsky91, Hirata99}:
\[
	p^+_A(j) \sim e^{-j/T}.
\]
However, as was first observed by \cite{Channon80}, such distributions for area-preserving maps with mixed regular and chaotic orbits appear to decay much more slowly:
\beq{ExitTimePowerLaw}
	p^+_A(j) \sim t^{-\gamma-1} .
\eeq
Such algebraic decay is a signal of the stickiness of stable structures in the phase space, see \Sec{Markov}.

The existence of the average \Eq{AverageExit} implies that $\gamma>0$. 
It can be shown that the exponent for $f$ and $f^{-1}$ are the same \cite{Easton93b}, thus \Eq{TransitTime} implies that the transit time distribution will have this exponent as well, and hence, by \Eq{ReturnDistribution}, so will the return distribution. Consequently, the cumulative recurrence distribution will decay as
\beq{PowerLaw}
	P^{return}_A(t) \sim t ^{-\gamma}. 
\eeq
Many studies have found $\gamma \simeq 1.5$ \cite{Venegeroles09}, though there are fluctuations that persist for large times \cite{Ceder13}, and it is still controversial whether the power law \Eq{PowerLaw} is valid asymptotically, and if it is, whether there are classes of dynamical systems for which the exponent $\gamma$ is universal.

\begin{prob}[Power Law Decay] Do exit and return-time distributions in typical volume-preserving maps with mixed regular and chaotic components have an asymptotic power law form? Is there a universal exponent $\gamma$?
\end{prob}

The exponent $\gamma$ is related to another measure of transport, the growth of the mean square displacement $\langle \|x_t-x_0\|^2\rangle$ with time. For Brownian motion, the squared displacement grows linearly in time and there exists a diffusion coefficient,
\[
	D = \lim_{T\to\infty} \frac{1}{2T}\langle \|x_T-x_0\|^2\rangle .
\]
However, it is observed that maps like \Eq{StdMap} sometime exhibits anomalous momentum diffusion with 
\beq{AnomalousDiffusion}
	\langle (y_t - y_0)^2\rangle \sim t^\beta .
\eeq
\textit{Super-diffusion}, i.e., $\beta >1$, is observed in the standard map when there are \textit{accelerator modes}, stable orbits that advance linearly in the momentum direction \cite{Chirikov79a, Karney82, RomKedar98, Venegeroles08a, Manos14, Albers14}. An accelerator island will drag nearby chaotic orbits along, so that even when the ensemble average does not include the accelerating island, the effect is that $\beta > 1$. The simplest accelerator mode forms at $k = 2\pi$, where after one step, $(x_1,y_1) = (x_0, y_0+1)$, however, there are probably high-period accelerator modes arbitrarily close to $k_{cr}(\phi)$. More general maps that do not have the vertical periodicity of the standard map do not have accelerator modes; however, the effect of sticky islands is still important.

Karney \cite{Karney83} showed that \Eq{PowerLaw} and \Eq{AnomalousDiffusion} can be related through the force correlation function, $C(t)$. Indeed the effective diffusion after $T$ steps is 
\[
	 \frac{1}{2T} \langle (y_T - y_0)^2\rangle = \tfrac12 C(0) + \sum_{t=1}^T C(t).
\]
Karney's result is that the power law \Eq{PowerLaw} implies $C(t) \sim t^{-\gamma +1}$, which then, by \Eq{AnomalousDiffusion}, gives
\beq{GammaBeta}
	\gamma + \beta = 3.
\eeq
Consequently, the algebraic decay of the recurrence distribution and anomalous diffusion are two sides of one coin. Since numerical experiments estimate that $\gamma \approx 1.5$, then \Eq{GammaBeta} implies that $\beta \approx 1.5$ as well. This has been confirmed by Venegeroles using the Perron-Frobenius operator to analytically estimate diffusion due to accelerator modes \cite{Venegeroles08a, Venegeroles09}.

Stickiness can also be studied using finite time Lyapunov exponents: the distribution of exponents is bimodal due to orbits sticking near elliptic regions \cite{Szezech05}. This is also observed in higher-dimensional systems \cite{Manchein14}.

\section{Markov Model for Transport}\label{sec:Markov}
The phase space of a typical area-preserving map is a complex mixture of islands and chaotic orbits, recall \Fig{StdMapLeaking}. Each island, a collection of nested invariant circles, is also surrounded by periodic orbits that have rational rotation numbers relative to the elliptic periodic orbit at the island's center. Each elliptic periodic orbit also generically forms islands. Numerically, there appear to be infinitely many islands comprising a \textit{fat fractal}: a chaotic region has nonzero area but its boundary has non-integer dimension \cite{Umberger85}. Thus, embedded in the chaotic sea is an infinite hierarchy of islands around islands, see \Fig{PhaseSpace7x7}.

A connected chaotic component can be partitioned into tree of states \cite{MacKay84, Meiss86a}. The nodes of the tree correspond to regions bounded by partial barriers. Each state is a portion of the chaotic component surrounding a given elliptic island, i.e., the accessible subset of a resonance zone. Transport between states is mediated by turnstiles. These could be lobes in resonance boundaries; however, at least for twist maps, there are also cantori between resonances (there are irrationals between every pair of rationals). Since lowest flux is that  through a cantorus, recall \Sec{CantoriBarriers}, these form the most restrictive partial barriers. The elliptic orbits inside a resonance zone also form islands that are separated by cantori. 

\InsertFig{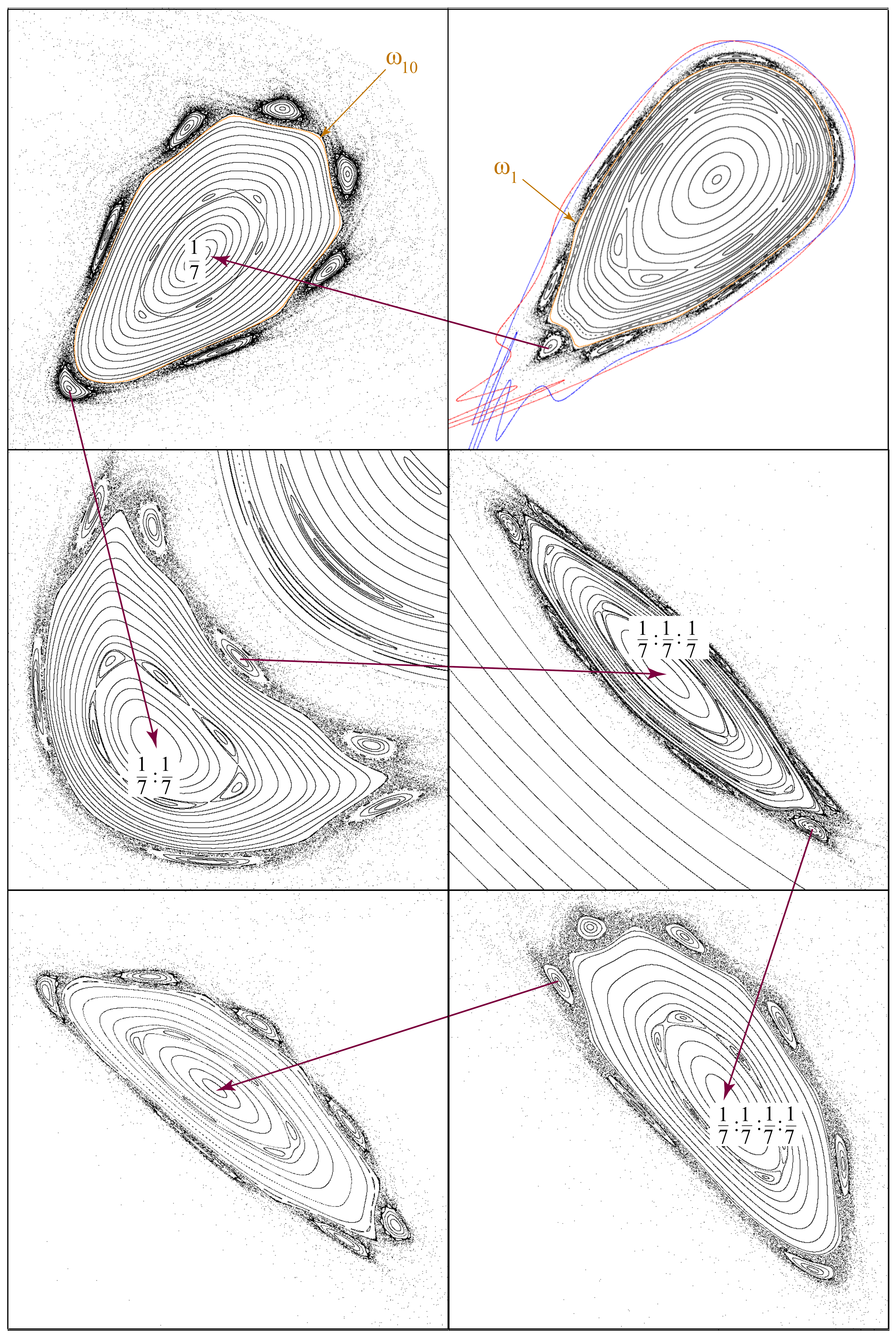}{Island around island structure for the H\'enon map \Eq{Henon} at $a=-0.17197997940$. Successive enlargements are indicated by the arrows. Boundary circles
for the states $S=1$ and $10$ are shown (brown). For this parameter, the map has a self-similar sequence of islands around islands, each with relative period $7$. Five levels of this sequence are shown. Reproduced with permission from Physical Review E 90(6):062923 (2014). Copyright 2014 AIP Publishing.}{PhaseSpace7x7}{0.7}

Transport through a chaotic region appears to be anomalous whenever it has a boundary component that is a stable invariant set. For area-preserving maps these are typically boundary circles of islands, the outermost of the encircling loops. It is these that give rise to the power law \Eq{ExitTimePowerLaw}.

The simplest model contains a single island so that the chaotic component is bounded by a single invariant circle. The stickiness of a boundary circle is caused by a sequence of cantori that limit on the boundary with ever-decreasing flux. As John Greene explained to us in 1983 \cite{MacKay84}, such a scenario yields a power law \Eq{ExitTimePowerLaw}; however, the exponent is much larger than that observed: $\gamma = 3.05$ \cite{Hanson85} or $\gamma = 3$ \cite{Chirikov99}. Alternatively, if only the hierarchy of islands around islands is kept, ignoring the cantori, the exponent is still too large: $\gamma = 2.25$ \cite{Zaslavsky97}. The insufficiency of either model was also confirmed by the observations of \cite{Weiss03}: a trajectory following either one of these hierarchies has an increasing probability of becoming trapped by secondary islands or by smaller flux cantori as time increases. This phenomena results in transient local density accumulations \cite{Smith87, Meiss94}. Note also that systems not smooth enough to have cantori and with no hierarchy of islands, such as some piecewise linear maps, have a different exponent, $\gamma = 2$ \cite{Altmann06}

The simplest partition that keeps both phenomena is a binary Cayley tree (nodes have a parent and two daughters), see \Fig{SelfSimilarTree}. Each node in the tree is labeled by a sequence $S = s_0s_1\ldots$, with $s_i \in \{0,1\}$. The root of the tree, denoted $\emptyset$, can be taken to be an \textit{absorbing state}: whenever a trajectory reaches this state, it is deemed to have escaped. For the H\'enon map of \Fig{PhaseSpace7x7}, the null state is the region outside the resonance zone formed by the stable and unstable manifolds of the hyperbolic fixed point (blue and red curves).

Transitions to the right and left on the tree correspond to distinct topological structures in phase space. A transition to the right corresponds to motion towards a given island boundary. The generically irrational rotation number of a boundary circle has a continued fraction expansion, and each of its convergents, \Eq{Convergents}, corresponds to a periodic orbit \cite{Greene86}. The resonance zones of every other convergent lie in the chaotic component, giving an infinite sequence of \textit{levels} that limit on the boundary circle. A transition from one level to the next corresponds to adding a ``1" to the state, $S \to S1$.

Each elliptic island within a state $S$ also has its own boundary circle and corresponding infinite sequence of levels of encircling periodic orbits. We denote the outermost of these new levels by $S0$; thus a transition to the left on the tree, $S \to S0$, corresponds to being trapped around an island of higher \textit{class} \cite{MacKay84, Meiss86b}. For example, in \Fig{PhaseSpace7x7}, the outermost elliptic islands in the upper right pane form a $(1,7)$ chain. 
This represents state $S=1$ in \Fig{SelfSimilarTree}. 
One island in this chain is enlarged in the upper left panel of \Fig{PhaseSpace7x7}; it is also encircled by a $(1,7)$ chain that has rotation number $\tfrac17$ relative to the seventh image, $f^7$, so that this orbit has period $49$ relative to $f$. In the figures this island is designated by its rotation number sequence $\tfrac17 : \tfrac17$. These are the outermost elliptic islands of \textit{class one} corresponding to the state $S=10$. 

If there is only one significant elliptic island in each state, then the tree is binary. More generally, some states in the tree can have more daughters when there are multiple islands in a given state, and some may have fewer if the periodic orbits are hyperbolic \cite{Alus14}.

\InsertFig{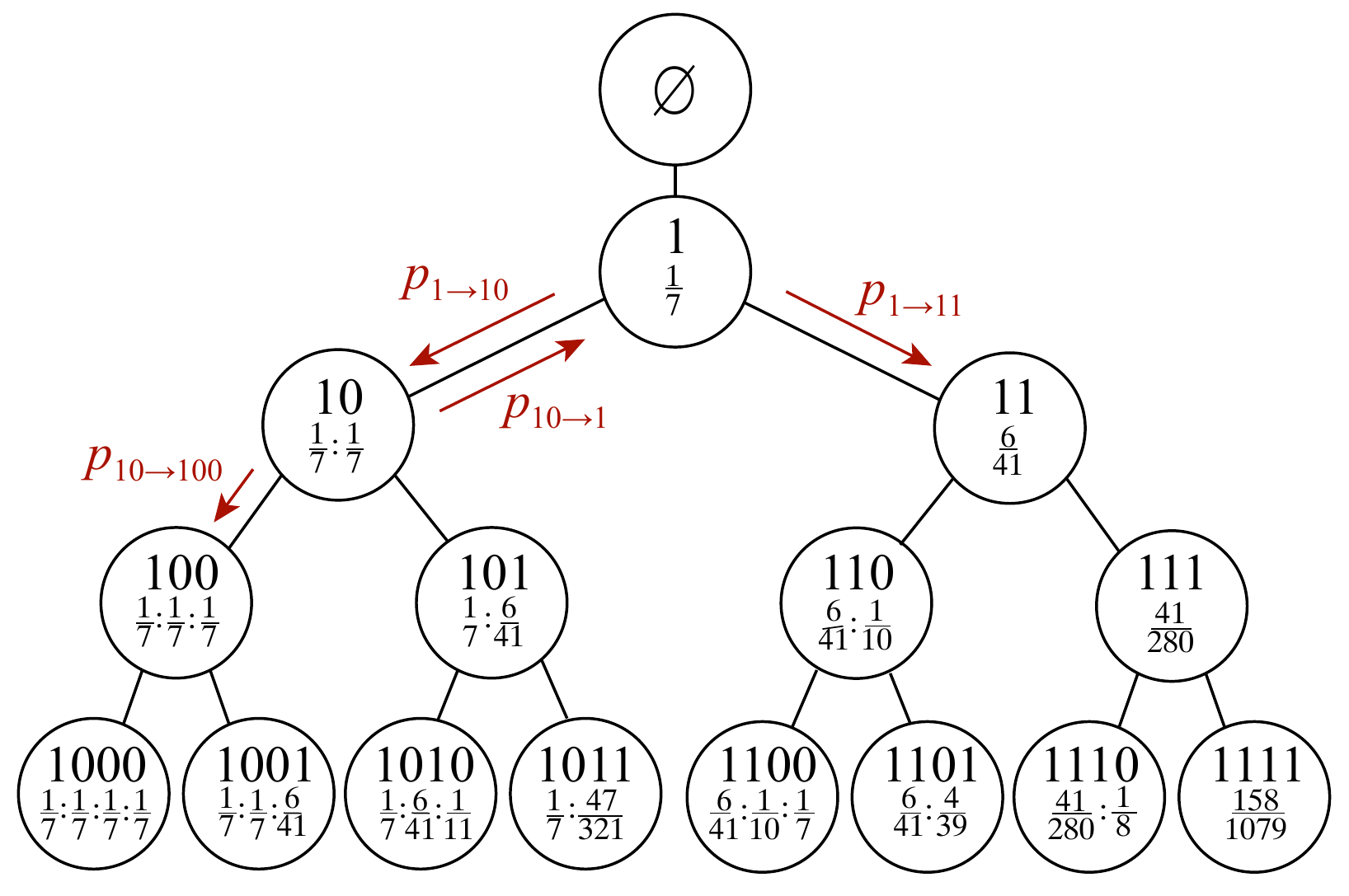}{Tree for the H\'enon map \Eq{Henon} at $a=-0.17197997940$ with the phase space shown in \Fig{PhaseSpace7x7} where it has a period-seven self-similar hierarchy, from \cite{Alus14}.
Each node is labeled by the state $S$ and the winding number sequence for the main island of that class, $\frac{p_1}{q_1}: \frac{p_2}{q_2} : \ldots$. Several transition probabilities, $p_{S \to S'}$, are also indicated.  Reproduced with permission from Physical Review E 90(6):062923 (2014). Copyright 2014 AIP Publishing.}{SelfSimilarTree}{0.6}

The simplest approximate model for transport on the tree of states is Markov. The assumption is that the chaos within a state $S$ causes rapid mixing, and so the probability of a trajectory leaving at any time is proportional to the flux, $\Delta W_{S,S'}$, of trajectories from $S$ to  state $S'$, where $S'$ is the parent, $DS$, or a daughter, $S0$ or $S1$. This flux is the area of the turnstile of the most resistant cantorus that divides the states and can be computed as an action difference, recall \Sec{Action}. Note that the flux is symmetric, $\Delta W_{S,S'}= \Delta W_{S',S}$, when the map is exact, recall \Sec{Exact}.
This model seems reasonable when the average exit time \Eq{AverageExit} is large compared to a Lyapunov time---the time for significant memory loss in the dynamics. However, for a near-integrable system, the mixing approximation probably will not work, and there are strong correlations that result in oscillatory exit time distributions \cite{RomKedar94}. 

\begin{prob}[Markov Transport Models]
Is there a limit in which the Markov model is valid? When can higher-order correlations be neglected?
\end{prob}

Thus a Markov model for transport on the tree is defined by transition probabilities
$p_{S\to S'}$ for each pair of connected nodes on the tree, recall \Fig{SelfSimilarTree}. Under the assumption that a trajectory has equal probability to be anywhere in $S$, the probability of such a transition is 
\beq{TransitionProb}
	p_{S \to S'} = \frac{\Delta W_{S,S'}}{A_S} ,
\eeq
where $A_S$ is the accessible area of $S$. By \Lem{ExitTime}, \Eq{TransitionProb} is exactly the inverse of the average transit time through the region.
Following \cite{Meiss86a}, it is convenient to categorize the change in transition probabilities on the tree by two ratios,
\beq{FluxRatios}
	w_{i}(S) = \frac{p_{S \to Si}}{p_{S \to DS}} , \quad
	\eps_{i}(S) = \frac{p_{Si \to S}}{p_{S\to DS}},
\eeq
for $i \in \{0,1\}$.
The first ratio measures asymmetry between motion ``down'' and ``up'' the tree, and the second measures the slowing of time scales deeper in the tree.

For example, renormalization theory implies that for a noble invariant circle, $w_1 = 0.053112$: it is $19$ times more likely for an orbit to move away from the circle---to the parent state---than to move one level deeper \cite{MacKay83}. For the self-similar sequence of period-seven islands shown in \Fig{PhaseSpace7x7}, the class ratio is even smaller, $w_2 = 0.014158$, implying that trapping around an island is an even rarer event \cite{Meiss86b}. For this reason, anomalous transport is observable only for extremely long trajectories. Accordingly, it is debatable whether an asymptotic regime is reached even for the longest computations (e.g., quadruple precision computations for orbits of length $10^{12}$ \cite{Weiss03}).

In the original Markov tree model, Ed Ott and I assumed that the tree is self-similar \cite{Meiss86a}.
This would asymptotically hold if every boundary circle were noble and every island chain had the same period relative to its parent. In this case the ratios \Eq{FluxRatios} are independent of the state $S$, though they depend on the choice of class, $i=0$, or level, $i = 1$. More generally, the results of \cite{Greene86} imply that the level hierarchy is approximately self-similar, and that for almost all boundary circles
\beq{LevelScaling}
	w_1 \simeq \eps_1^{3.05} .
\eeq
Similarly, there are cases in which the class hierarchy is asymptotically self-similar \cite{Meiss86b}. Numerical studies again show that the scaling coefficients are related, 
\beq{ClassScaling}
	w_0 \simeq \eps_0^{2.19} ,
\eeq
even as the period of the successive islands varies.

If the binary tree is exactly self-similar, then there is an integral equation for the recurrence distribution \cite{Meiss86a}. A solution to this gives the power law \Eq{ExitTimePowerLaw}, with an exponent determined by the dispersion relation
\[
	w_0 \eps_0^{-\gamma} + w_1 \eps_1^{-\gamma} = 1 .
\]
Given the relations, \Eq{LevelScaling}-\Eq{ClassScaling}, there are still two parameters needed to solve for $\gamma$. Reasonable values for the $\eps_i$ give $\gamma = 1.96$, which is closer to the observed value, $\gamma \simeq 1.5$, than the models with no tree hierarchy. The analysis of \cite{Meiss86a} implies that each branch added to the tree decreases the value of $\gamma$.

Note, however, that for a typical map, neither self-similar scenario will hold, and the scaling factors \Eq{FluxRatios} will depend upon the state $S$. Recent studies have modeled the lack of self-similarity statistically. The idea is that an orbit will sample different regions of phase space, and its long-time behavior will be equivalent to an ensemble of scalings. Effectively, the coefficients of the Markov tree model should be taken as a random draw from some probability distribution. Numerical studies do appear yield different values of $\gamma$ in different regions of phase space \cite{Weiss03, Abud13}. Using such a model, Cristadoro and Ketzmerick find $\gamma = 1.57$, in remarkable agreement with numerical experiments \cite{Cristadoro08}. A similar model, using a different ensemble was studied in \cite{Ceder13}. Indeed, it is not clear which random ensemble of scaling coefficients is valid, though there may be some sense in which there is a universal ensemble \cite{Alus14}.
The ensemble picture also provides an explanation for the difficulty of observing an asymptotic regime: the correlation function of the fluctuations in the mean decay exponent have a log-periodic behavior that decays very slowly in time \cite{Ceder13}.

The Markov model, then, not only gives a qualitative explanation for the power law decay  of the exit time distribution \Eq{ExitTimePowerLaw}, but also gives a value of $\gamma$ close to that observed. Of course, this is only for the area-preserving case.

\begin{prob}[Higher-Dimensional Transport] Are there Markov tree models that quantitatively explain transport in higher-dimensional volume-preserving and symplectic maps?
\end{prob}

\section{The Next Thirty Years}\label{sec:Conclusion}

I have tried to summarize both the state of the art in the theory of transport for conservative dynamical systems and to indicate some directions that seem to me to be still ripe for future research. 

There are other fruitful techniques that I did not include in this survey, such as the characterization of mixing in two-dimensional flows by the braiding of trajectories \cite{Boyland00, Finn11}, and the set theoretic methods based on the Perron-Frobenius or Koopman operator formalisms \cite{Froyland99, Froyland05, Budisic12, Bollt13}. I also did not discuss the continuous time and space approximation of anomalous transport models to obtain phenomenological, fractional kinetic equations. This theory, which is reviewed in \cite{Zaslavsky97, Zaslavsky02}, does not have as close a connection with the underlying topology of the dynamics as the discrete Markov tree.
Finally, I did not discuss the application of the ideas of turnstiles to semi-classical and quantum mechanics. Indeed, in \cite{MacKay84} we speculated
that classical cantori whose turnstiles were small in units of Planck's constant might play a significant role in the semiclassical mechanics of molecules, and MacKay and I later applied this idea to compute the ionization threshold for hydrogen in a strong microwave field \cite{MacKay88a}.
Several recent advances in this area confirm our original speculation \cite{Michler12, Richter14}.

There is much that remains intriguing about this problem: the infinite complexity of phase space and the long-time correlations that result in apparent power law decays and anomalous diffusion. Even after thirty years, I still cannot resist the allure of phase space portraits for simple area-preserving maps. 

In \cite{MacKay84} we noted
\begin{quote}
The theory is not yet complete, but we believe that the ideas introduced here are of central importance for a broad range of systems that are neither completely ordered nor completely chaotic. 
\end{quote}

\noindent
While there has been considerable progress, I think---as indicated by the series of questions posed in this paper---that there is still much to do in the study of transport.

\begin{acknowledgments}
The work was partially supported NSF grant DMS-1211350.
I would like to thank Robert MacKay, Ian Percival and Bob Easton for long and fruitful collaborations, many discussions, and a profusion of insights.
\end{acknowledgments}
\bibliographystyle{alpha}
\bibliography{ThirtyYears}

\end{document}